\documentclass[aps,pre,reprint]{revtex4-1}
\usepackage[utf8]{inputenc}
\usepackage{textcomp}
\usepackage{color,xspace}
\usepackage{graphicx}
\usepackage{upgreek}
\usepackage{url}
\usepackage{hyperref}

\renewcommand{\BibitemShut}[1]{}
\begin{document}
\title{Configurational temperature in dusty plasmas}
\date{\today}
\author{Michael Himpel}
\email{himpel@physik.uni-greifswald.de}
\affiliation{Institute of Physics, University Greifswald, Felix-Hausdorff-Straße 6, 17489 Greifswald, Germany}
\author{Andr\'e Melzer}
\affiliation{Institute of Physics, University Greifswald, Felix-Hausdorff-Straße 6, 17489 Greifswald, Germany}

\begin{abstract}
The temperature of a dust ensemble in a dusty plasma is one of its most fundamental properties. Here, we present experiments using the configurational temperature as a tool for the temperature analysis in dusty plasmas. Using a model of the particle interactions, the configurational temperature allows us to determine the temperature of the dust ensemble from measurements of the particle positions, rather than particle velocities. The basic concept will be presented and the technique is applied to two-dimensional finite clusters as well as three-dimensional data from an extended dust cloud. Additionally the configurational temperature can be used to derive the particle charge and the screening length from a comparison with the standard kinetic temperature.
\end{abstract}

\maketitle

\section{Introduction}
A dusty plasma often is generated by injecting spherical microparticles into a plasma. Due to the high mobility of the electrons, the micrometer-sized particles (called \emph{dust}) typically attain a negative charge of several thousands of elementary charges. On the one hand, the micron-sized particles are large enough to be directly observable with video cameras when the particles are illuminated by a laser. On the other hand, the particles are small enough to exhibit measurable Brownian motion\,\cite{QuinnPRE,NunomuraPRL,schmidt_heating}. This stochastic motion of the particles, which mostly results from collisions with the neutral gas, can be associated with a temperature. To clarify, here and in the following the term temperature is associated with the random motion of the particles and not the surface or bulk temperature of the particle material itself. This temperature is an important basic property of particle systems in a plasma. It can be used, for example, to relate the dusty plasma with thermodynamic properties\,\cite{heat,NosenkoPOP13,SchablinskiPOP19} or to characterize crystaline states\,\cite{ThomasNature,MelzerPRE53,Hayashi_1994,ChuPRL72,crystall, DietzPRE} of the ensemble.

The temperature of such particles in a dusty plasma is usually determined experimentally by measuring and analyzing the velocity of individual particles\,\cite{schmidt_heating,LiuPRE78} or by particle image velocimetry\,\cite{stereopiv, tomopiv}. The kinetic dust temperature $T$ is then derived from the root mean square of the particle velocities $v$ as 
\begin{equation}\label{eq_intr}
\frac{1}{2} k_\mathrm{B} T = \frac{1}{2} m_\mathrm{d}\langle v^2 \rangle
\end{equation}
with $m_\mathrm{d}$ being the particle mass and $k_\mathrm{B}$ being Boltzmann's constant.

In this paper, we present an alternative method to determine the temperature of a dusty plasma. This concept, called \emph{configurational temperature}, is adopted from analyses in the fluid community\,\cite{confTempSuspension, fluid2}. There, the random deviations of the particles from their equilibrium positions are used as a measure of temperature. The local excursions of a particle in the local confinement reflects its thermal agitation. Thereby, the forces from neighboring particles determine the local confinement.

The important difference between configurational temperature and kinetic temperature for the experimenter is the fact, that the configurational temperature relies on the measurement of the particle positions instead of their velocities. Velocity measurements using video cameras can be often problematic, as a sufficiently high framerate has to be achieved which competes with the low-light conditions that are usual in laboratory imaging situations\,\cite{SchmidtPOP23,schmidt_heating}. Moreover, tracking the particles from one frame to the next is not required using the configurational temperature. The possibility to measure properties of the system by analyzing  images of the stochastic particle positions allow the use of a less sophisticated and thus less expensive hardware.

Furthermore, a comparison between kinetic and configurational temperature allows to determine the particle interaction e.g. in terms of the dust charge since  the configurational temperature makes use of the interparticle forces. Thus, a model of the particle interaction is necessary for the derivation of the configurational temperature.

Here, we will describe the concept of configurational temperature for the special conditions in dusty plasmas. Then we present its application to two-dimensional simulated data, two-dimensional laboratory data and to three-dimensional data. Additionally we propose an approach to extract the particle charge from the comparison of the configurational and kinetic temperature.

\section{Temperature definitions}
We will start with a general temperature definition. From that, as special cases, the kinetic and configurational temperature are derived.

\subsection{General formulation}
The most general definition of a temperature in a classical $N$-particle ensemble is given by\,\cite{theo_CT1, theo_CT2}
\begin{equation}\label{eq_general}
k_\mathrm{B} T = \frac{\langle \nabla \mathcal{H}(\mathbf{\Gamma}) \cdot \mathbf{B}(\mathbf{\Gamma})  \rangle}{ \langle \nabla \cdot \mathbf{B}(\mathbf{\Gamma})  \rangle  }.
\end{equation} 
There, $\mathcal{H}$ is the system's Hamiltonian which is the sum of the kinetic energy $K(\mathbf{p}_i)$ and the conservative $N$-body potential $V(\mathbf{q}_i)$ where $\mathbf{q}_i$ are the $N$ generalized coordinate vectors and $\mathbf{p}_i$ are the $N$ conjugate momentum vectors. The angled brackets indicate the ensemble average. The full phase space is then represented by the coordinate set $\mathbf{\Gamma} = [ \mathbf{q}_1, \dots, \mathbf{q}_N, \mathbf{p}_1, \dots, \mathbf{p}_N]$. $\mathbf{B}(\mathbf{\Gamma})$ is any continuous and differentiable function in phase space. Different temperature definitions now arise from differently chosen $\mathbf{B}(\mathbf{\Gamma})$\,\cite{theo_CT2}.

Usually, the kinetic temperature of dust particles in a plasma is measured by analyzing the particle velocities. This approach is equivalent to choose the phase space function as $\mathbf{B}(\mathbf{\Gamma}) = [ 0, \dots, 0, \mathbf{p}_1, \dots, \mathbf{p}_N]$. Equation\,(\ref{eq_general}) considering a $N$-particle system with particles of mass $m$ then yields
\begin{eqnarray}\label{eq_mv}
& & \frac{\left\langle \frac{1}{N}\sum_{i=1}^{N} \left(\nabla \cdot \mathcal{H}\right) \mathbf{p}_i  \right\rangle}{ \langle  \frac{1}{N}\sum_{i=1}^{N} \nabla \cdot \mathbf{p}_i  \rangle  } \nonumber \\ 
& = &  \left\langle\frac{1}{N} \sum_{i=1}^{N} \frac{\mathbf{p}_i^2}{m}\right\rangle = D k_\mathrm{B} T_\mathrm{kin} 
\end{eqnarray}
which is an analogous formulation of the $N$-body equipartition theorem with $D$ being the dimensionality of the system. According to this definition, the temperature $T_\mathrm{kin}$ is a measure of the mean kinetic energy of the particles in the system and hence a function of the particle velocity or momentum.

In an experiment to measure the kinetic temperature of a particle system, one can either directly determine the root-mean square of the measured particle velocities as suggested by Eq.\,(\ref{eq_intr}). Usually it is more robust to determine the velocity distribution function and then fit this distribution with a Maxwell-Boltzmann distribution using $T_\mathrm{kin}$ as a free parameter\,\cite{schmidt_heating}. In this paper, the latter method has been used as its results are less influenced by measurement noise.

\subsection{Configurational temperature}
By choosing a proper $\mathbf{B}(\mathbf{\Gamma})$, the general temperature definition in Eq.\,(\ref{eq_general}) can be decoupled from the momentum terms in $\mathbf{\Gamma}$. This is done by setting $\mathbf{B}(\mathbf{\Gamma}) = -\nabla V(\mathbf{q}_i)$. The resulting configurational  temperature
\begin{equation}\label{eq_confT1}
k_\mathrm{B} T_\mathrm{conf} = \frac{-\left\langle \nabla V(\mathbf{q}_i) \nabla V(\mathbf{q}_i) \right\rangle}{\left\langle \nabla^2 V(\mathbf{q}_i) \right\rangle}
\end{equation}
depends only on the particle positions $\mathbf{q}_i$ and not on the particle's momenta $\mathbf{p}_i$. In this definition, one interprets the temperature of a particle as a relation between thermal agitation and the net restoring forces on a particle. Particles at a low temperature will on average be situated close to the force equilibrium position most of the time, whereas particles with a higher temperature can ``climb up'' higher along the force gradient and thus be on average situated further away from the equilibrium position. The ensemble average of the excursion widths from equilibrium then yields the temperature. This technique has been successfully applied to colloidal suspensions\,\cite{confTempSuspension} and is used in molecular dynamics simulations\,\cite{CTmd}. We now like to address the suitability of the configurational temperature for dusty plasmas. In an experiment, this definition allows a temperature measurement by using snapshots of particle positions in a dusty plasma. This might simplify temperature measurements in experiments where it is difficult to obtain accurate velocity distributions. Further, a tracking of particles is not required. However, the potential $V$ must be known or assumed.

\subsection{Force model for Dusty Plasmas}
Since $\mathbf{F}_j = -\nabla V(\mathbf{q}_j)$, Eq.\,(\ref{eq_confT1}) can then be written as
\begin{equation}\label{eq_confT}
k_\mathrm{B} T_\mathrm{conf} = -\frac{\left\langle \sum_{j=1}^N \mathbf{F}_j^2 \right\rangle}{\left\langle\sum_{j=1}^N  \nabla_j \cdot \mathbf{F}_j \right\rangle}.
\end{equation}
The particles are subject to two governing forces: The electrostatic particle-particle interaction force $\mathbf{F}^\mathrm{el}$ and the confinement force $\mathbf{F}^\mathrm{c}$. To express Eq.\,(\ref{eq_confT}) with the actual forces acting on the particles one needs to express the force itself and its divergence. Microparticles that are injected into a plasma environment usually attain a highly negative charge. The free electrons and ions result in a shielded pair-wise repulsive interaction between the dust particles. The particle interaction can then be modelled by a shielded Coulomb interaction\,\cite{KonopkaPRL84,WangPRL86,LaiPRE60} as
\begin{equation}\label{eq_force}
\mathbf{F}_{ij}^\mathrm{el} = \frac{Q_i  Q_j}{ 4\pi\epsilon_0  } \left( \frac{1}{r_{ij}^2} + \frac{1}{r_{ij} \lambda_\mathrm{s} } \right) e^{-r_{ij} / \lambda_\mathrm{s}} \frac{ (\mathbf{q}_i-\mathbf{q}_j)}{r_{ij}}
\end{equation}
with $r_{ij} = || \mathbf{q}_i-\mathbf{q}_j||$ being the euclidean distance between particle $i$ and $j$ and $\lambda_\mathrm{s}$ being the effective shielding length. The total interaction force acting on each single particle $j$ is then generated by all particles $i$ yielding
\begin{equation}\label{eq_forceSum}
\mathbf{F}_j^\mathrm{el} = \sum_{i \neq j}^N \frac{Q_i  Q_j}{ 4\pi\epsilon_0  } \left( \frac{1}{r_{ij}^2} + \frac{1}{r_{ij} \lambda_\mathrm{s} } \right) e^{-r_{ij} / \lambda_\mathrm{s}} \frac{ (\mathbf{q}_i-\mathbf{q}_j)}{r_{ij}}
\end{equation}
The divergence $\nabla_j \cdot \mathbf{F}_j^\mathrm{el}$ can then directly be calculated as
\begin{eqnarray}\label{eq_divSum}
\nabla_j \cdot \mathbf{F}_j^\mathrm{el} &=&  \sum_{i \neq j}  \left[  \frac{Q_i  Q_j}{ 4\pi\epsilon_0  } \left( \frac{1}{r_{ij}^3} + \frac{1}{r_{ij}^2 \lambda_\mathrm{s} } + \frac{1}{r_{ij}  \lambda_\mathrm{s}^2 } \right) \times \right. \\ \nonumber
&& \left. \times e^{-r_{ij} / \lambda_\mathrm{s}}   \right].
\end{eqnarray}
with $\mathbf{1}$ denoting the unity vector.

One should note that the particle interaction model neglects ion drift motion. Such an ion drift motion influences the shape of the particle potential and hence the interaction forces. Simulations show, that a significant influence of streaming ions can be found with Mach-numbers $M>0.1$ \cite{wakePRE}. Then, wake-field effects might come into play. For our 3D experiments we assume an isotropic interaction neglecting the influence of wake fields. For the 2D case, a wakefield is present in the vertical direction. However, we are interested only in the horizontal interaction between the dust where a shielded Coulomb interaction is applicable \cite{KonopkaPRL84}.

The confinement force can be assumed to result from a harmonic confinement potential and thus to be linear as $\mathbf{F}^\mathrm{c}_j = -k \mathbf{q}_j$ with $k$ being the confinement strength for the specific experimental conditions\,\cite{LaiPRE60,KonopkaPRL84}. The divergence of this confinement force is then simply $\nabla \cdot \mathbf{F}^\mathrm{c}_j = -D k$ with $D$ denoting the dimensionality of the system.

Finally, to express the configurational temperature with the forces acting on a particle cluster in a dusty plasma, this can be combined to
\begin{equation}\label{eq_CTfinal}
k_\mathrm{B} T_\mathrm{conf} = -\frac{\left\langle \sum_{j=1}^N  \left(\mathbf{F}_j^\mathrm{el} + \mathbf{F}^\mathrm{c}_j \right)^2 \right\rangle}{\left\langle\sum_{j=1}^N \left(  \nabla_j \cdot \mathbf{F}_j^\mathrm{el} \right) + N D k    \right\rangle}.
\end{equation}

\subsection{Normal Mode Analysis}
To be able to compare the configurational temperature results with an established method, the experimental data for 2D systems has also been investigated using normal mode analysis\,\cite{Melzer03,Schweigert95c,Nelissen06}. There, a two-dimensional cluster of dust particles is characterized by its
total energy
\begin{equation}\label{nma1}
E=\frac{1}{2}k\sum\limits_{j=1}^N r_j^2 +
\frac{Q^2}{4\pi\epsilon_0}\sum\limits_{i>j}\frac{1}{r_{ij}}e^{-r_{ij}/\lambda_{\rm
s}}.
\end{equation}
It should be noted, that the potential energy of particle $j$ is assumed to be harmonic ($E_\mathrm{pot}=(1/2) k \mathbf{r}_j^2$), just as in the configurational temperature approach in the previous section. Using the normalizations for distances \,\cite{Schweigert95c}
\begin{equation}\label{nma2}
r_0 = \left[\frac{Q^2}{4\pi\epsilon_0}\frac{2}{k}\right]^{1/3}
\end{equation}
and energies
\begin{equation}\label{nma3}
E_0 = \left[\left(\frac{Q^2}{4\pi\epsilon_0}\right)^2\frac{k}{2}\right]^{1/3}
\end{equation}
the total energy is given as
\begin{equation}
E=\sum\limits_{j=1}^N r_j^2 +
\sum\limits_{i>j}\frac{1}{r_{ij}}e^{-r_{ij}\kappa}.
\end{equation}
with the screening strength $\kappa=r_0/\lambda_{\rm s}.$ Hence, it is
seen that the dynamics of such a cluster only depends on particle number $N$ and screening strength $\kappa$.

The dynamics of such a cluster is described in terms of its normal modes \,\cite{Schweigert95c,Nelissen06}. The normal modes are derived from the dynamical matrix (which contains the second derivative of the total energy with respect to all particles and coordinates). The eigen vectors of the dynamical matrix describe the mode oscillation patterns and the eigen values their oscillation frequency. There are $2N$ eigen modes for a 2D system of $N$ particles.

Now, the experimental thermal Brownian motion of the particles with
velocity $\mathbf{v}_j(t)$ is
decomposed into their respective contributions to the eigen modes as
\begin{equation}
    v_{\ell}(t) = \sum\limits_{j=1}^N \mathbf{v}_j(t)\cdot \mathbf{e}_{\ell,j} \quad,
\end{equation}
where $\mathbf{e}_{\ell,j}$ is the eigen vector for particle $j$ in mode
number $\ell$. Then, finally,
the normal mode spectrum of each mode is calculated, yielding the
spectral power density as
the Fourier transform of $v_{\ell}(t)$ as
\begin{equation}
    S_{\ell}(\omega)= \frac{2}{T}\left|\int\limits_{-T/2}^{T/2}
v_{\ell}(t) e^{i\omega t} dt\right|^2 \quad.
\end{equation}
The experimental mode spectrum is then compared with the theoretical
mode spectrum, i.e. the
eigen values of the dynamical matrix \,\cite{Melzer03}. As mentioned
above, the theoretical
spectrum only depends on $\kappa$ (for a known particle number $N$). For
the comparison
with the measured mode spectrum a value of $\kappa$ (or equivalently
$\lambda_{\rm s}$) is
prescribed and from the absolute size of the cluster and its spectrum then $r_0$ and $E_0$ are derived (note,
that only these two
values describe the behavior of all $2N$ modes). From the knowledge of
$r_0$ and $E_0$ the
particle charge $Q$ is extracted. This way $\lambda_{\rm s}$ - $Q$
-pairs are derived that
yield the observed mode spectrum, see also Ref.~\,\cite{Melzer03}. Hence, from the dynamical and structural properties of such a cluster the defining quantities $\lambda_\mathrm{s}$ and corresponding $Q$ can be derived.

Additionally, a mode temperature for mode $\ell$ can be defined from the
spectral power density by observing that
\begin{equation}
    \langle v_{\ell}^2\rangle = \int\limits_0^\infty S_{\ell} (\omega)
d\omega \quad.
\end{equation}
The mode temperature $T_{\ell}$ is then just given as 
\begin{equation}\label{eq_mode}
\frac{1}{2} m\langle v_{\ell}^2\rangle = \frac{1}{2} k_\mathrm{B} T_{\ell}.
\end{equation}
The overall mode temperature can be calculated as the mean over all mode
temperatures. In the
absence of measurement errors, the mean mode temperature coincides with
the mean kinetic
temperature, since they are both derived from averages over the particle velocities.

\section{Experiment description}
\begin{figure}
\includegraphics[width=\columnwidth]{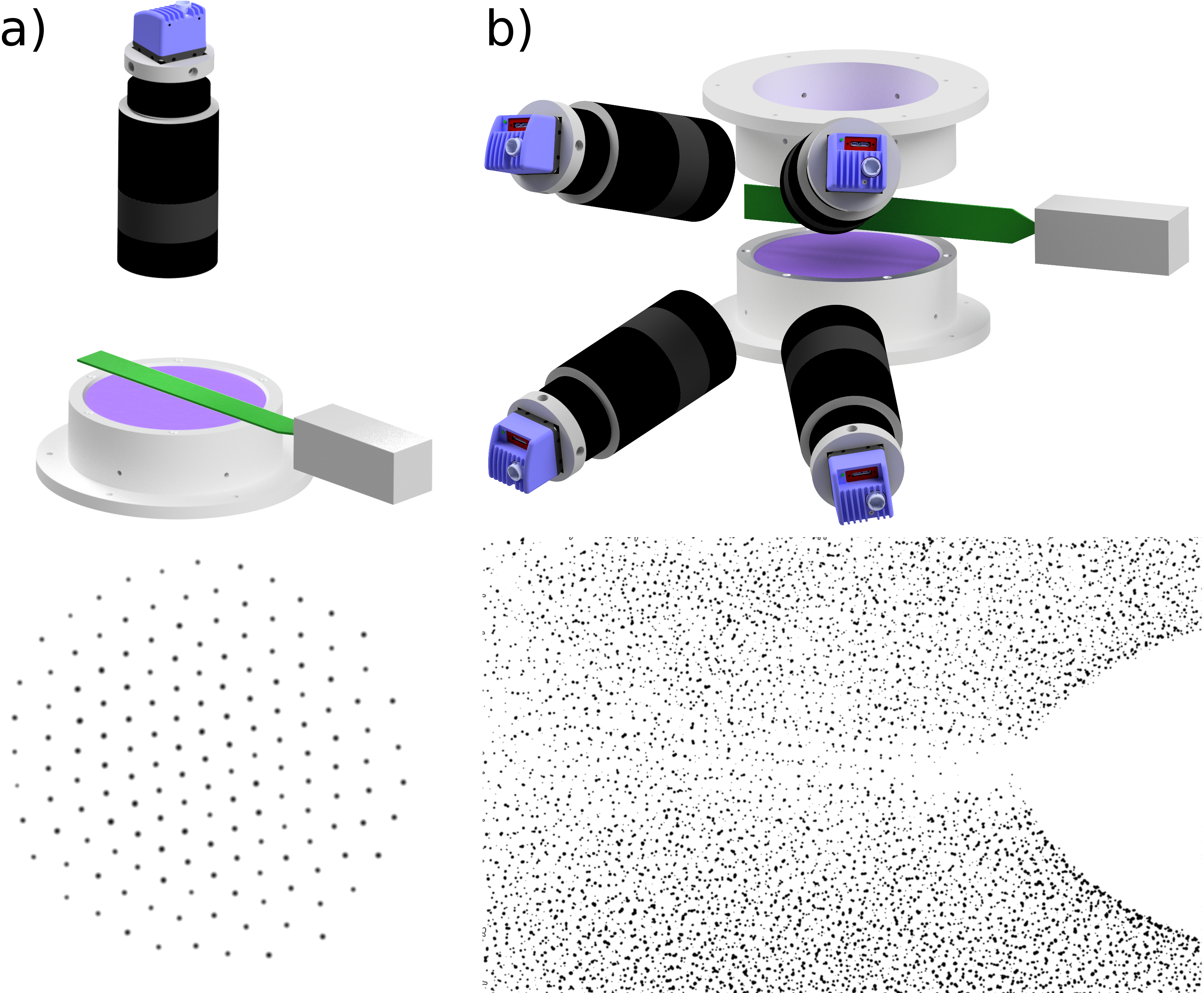}
\caption{(a) Experiment to confine a flat microparticle structure under gravity conditions. The field-of-view from a top-view camera covers all particles. An inverted measurement image is shown below the setup. (b) Symmetric discharge to confine a three-dimensional dust cloud under microgravity conditions. Four cameras observe a common field-of-view to retrieve the three-dimensional particle positions. An inverted measurement image is shown below the setup.}
\label{fig_experiment}
\end{figure}
The configurational temperature has been derived in two different dusty plasma systems: two-dimensional finite clusters\,\cite{JuanPRE58,KlindworthPRB61,Cheung_2003,Melzer03}, and three-dimensional extended dust clouds. The measurements of two-dimensional clusters have been performed in a laboratory setup as sketched in Fig.\,\ref{fig_experiment}(a). A capacitively coupled radio frequency (rf) plasma is created in a vacuum chamber using Argon gas at pressures between 4 and $9\,\mathrm{Pa}$. The variation of the gas pressure is known to result in a change of the particle temperature and hence allows to investigate clusters with different temperatures\,\cite{schmidt_heating}. Here, the rf-power was set to $8\,\mathrm{W}$. Particles of $7\,\upmu\mathrm{m}$ diameter have been dropped into the plasma. They are trapped in the sheath by a force balance of gravitational and electrostatic fields. The lower rf  electrode has a parabolic depression that results in a (harmonic) horizontal particle confinement. The particles are illuminated by a thin laser sheet. The observation camera is equipped with a sCMOS-sensor. The particle cluster is recorded at a framerate of $160\,\mathrm{fps}$ with a spatial resolution of approximately $10\mathrm{\upmu m / px}$. The particle positions in the images are determined up to the subpixel-level using the known Gaussian filter algorithm\,\cite{crocker, feng, ivanov_heating}. 

The three-dimensional dust cloud has been investigated on parabolic flights that allow to generate large, three-dimensional particle systems\,\cite{3D_MenzelPRL,himpel18,Thomas_PFC2001}. There, a symmetric parallel-plate discharge has been operated in push-pull mode. A laser sheet with a width of approximately $2\,\mathrm{mm}$ illuminates a slice through the center of the dust cloud as shown in Fig.\,\ref{fig_experiment}(b). The scattered light from the particles is then captured with a stereoscopic four-camera system\,\cite{himpel18}. From that, the three-dimensional trajectories of particles in the illuminated volume are reconstructed; see Ref.\,\cite{himpel18} for details.

To test the different temperature definitions also on data with known parameters, a two-dimensional dust cluster comparable to the measured one was simulated using molecular dynamics (MD). In this simulation, the particles interact via a screened Coulomb interaction with a screening length $\lambda_\mathrm{s}=500\,\mathrm{\upmu m}$ and a particle charge of $Q=12100e$ with $e$ being the electron charge. The confinement force is set to $\mathbf{F}^\mathrm{c}_i = -k  \mathbf{q}_i = -\left(2.2\cdot 10^{-11}\,\mathrm{kg/s^2}\right) \mathbf{q}_i$. This resulted in a cluster with a mean particle distance of $b=483\,\upmu\mathrm{m}$ similar to the experiment. The cluster consists of $500\,\mathrm{particles}$. The temperature of the particles was set to $800\,\mathrm{K}$
using a Langevin thermostat\,\cite{simu_book}. For a realistic comparison, we have chosen absolute parameters that are of the same order as those expected for the experiments.

\section{Results}
Here, now the results of the MD model and the experimental data are described.

\subsection{Molecular-dynamics simulation}
\begin{figure}
\includegraphics[width=\columnwidth]{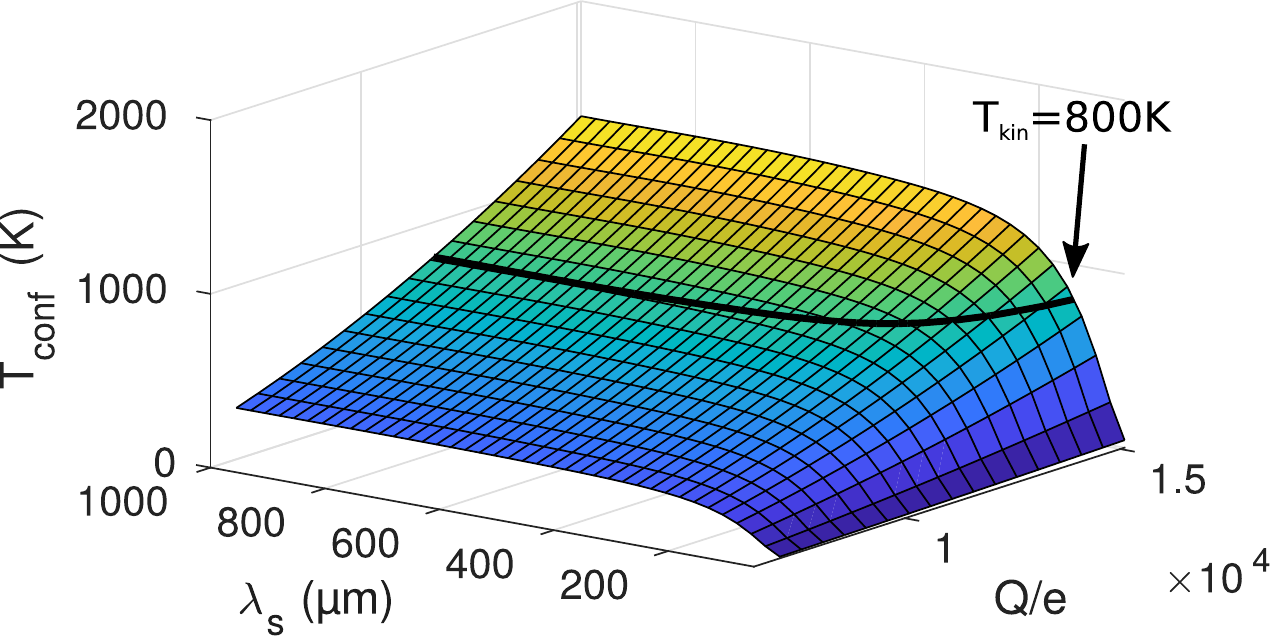}
\caption{Configurational temperature map over a grid of different particle charges $Q$ and screening lengths $\lambda_\mathrm{s}$. The solid black line shows the contour of the surface at the kinetic temperature $T = 800\,\mathrm{K}$ used in the simulation.}
\label{fig_surf}
\end{figure}
\begin{figure}
\includegraphics[width=\columnwidth]{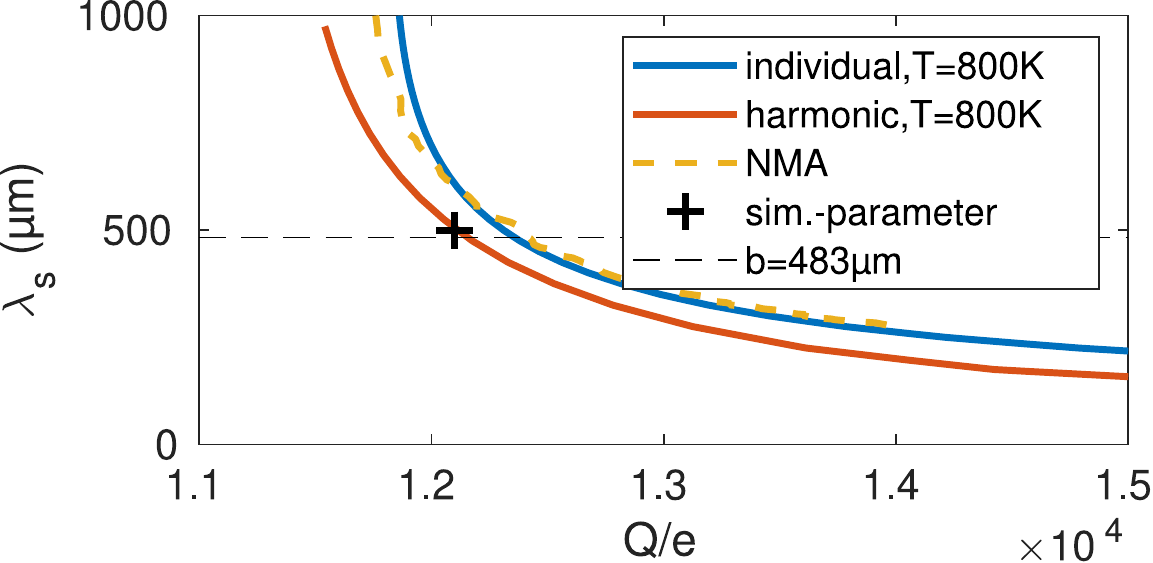}
\caption{The plot shows curves in the $Q$-$\lambda_\mathrm{s}$-plane that yield the parameters of the simulated dust cluster. The dashed line represents the results from the mode analysis, the two solid lines are obtained by setting $T_\mathrm{conf}^\mathrm{ind}(Q, \lambda)=800\,\mathrm{K}$ and $T_\mathrm{conf}^\mathrm{c}(Q, \lambda)=800\,\mathrm{K}$. The crossmark is placed at the actual simulation parameter position ($\lambda_\mathrm{s}=500\,\mathrm{\upmu m}$ and $Q=12100e$). The dashed horizontal line indicates the mean interparticle distance. }
\label{fig_methods}
\end{figure}
Using the simulated particle data, we will first study the different temperature definitions. The simulated particle velocities closely follow a Maxwell--Boltzmann distribution. From that, the kinetic temperature  can be determined using Eq.\,(\ref{eq_intr}) within a 95\% confidence interval as $T_\mathrm{kin} = 804\pm 4 \,\mathrm{K}$ in close agreement with the temperature $T=800\,\mathrm{K}$ set in the simulation. The mean mode temperature from NMA according to Eq.\,(\ref{eq_mode}) is found to be $T_\mathrm{mode} = 801\,\mathrm{K}$ again in agreement with the simulation parameters. Using the known parameters for $Q$, $k$ and $\lambda_\mathrm{s}$ from the simulation, the configurational temperature is determined as $T_\mathrm{conf} = 843\,\mathrm{K}$ by using Eq.\,(\ref{eq_CTfinal}). This is only slightly larger than the prescribed value.

The configurational temperatures could be computed here because the particle charge $Q$ and the screening length $\lambda_\mathrm{s}$ are known parameters from the MD simulation. But generally, the particle charge and the screening length are not known a priori. In that  case, one can try to derive these values from a comparison of configurational and kinetic temperature. Therefore, one can compute the configurational temperature map over a relevant grid of parameters $Q$ and $\lambda_\mathrm{s}$. Such a computation is shown in Fig.\,\ref{fig_surf}. The configurational temperature surface at various values of particle charge and screening length parameter is determined. This configurational temperature is compared with the kinetic temperature of $T=800\,\mathrm{K}$ that is indicated by the black line. From that, possible $\lambda_\mathrm{s}(Q)$ values are found that result in a configurational temperature identical to the kinetic temperature. Figure\,\ref{fig_methods} shows this $\lambda_\mathrm{s}(Q)$ relation. For comparison, the dashed line shows the $\lambda_\mathrm{s}(Q)$-function that has been found from the mode analysis [Eq.\,(\ref{eq_mode})]. It can be said, that all curves come very close to the preset simulation parameters (cross). When the estimation of either the charge $Q$ or the screening length $\lambda_\mathrm{s}$ is possible, the remaining parameter can be extracted from this curve. It should be noted that the $\lambda_\mathrm{s}(Q)$ curve is very steep in the region close to the actual simulation parameter. Thus, even a guessed screening length with large tolerance results in a quite accurate particle charge estimate. For example, taking $\lambda_\mathrm{s}$ as the interparticle distance $b=483\,\upmu\mathrm{m}$ one yields a particle charge $Q=11800e$ compared to the actual value of $12100e$. Many previous experiments\,\cite{HomannPRE56,LiuPRE71,HomanPLA,nunomura} found that the screening strength $\kappa = b/\lambda_\mathrm{s} \approx 1$, so that $\lambda_\mathrm{s}\approx b$ is a reasonable choice. However, $\kappa$ may be different in other experiments, where strong forces (i.e. the gravitational force) result in particles that are located considerably closer to eachother than their corresponding shielding length.

\subsection{Analysis of confinement forces}
\begin{figure}
\includegraphics[width=\columnwidth]{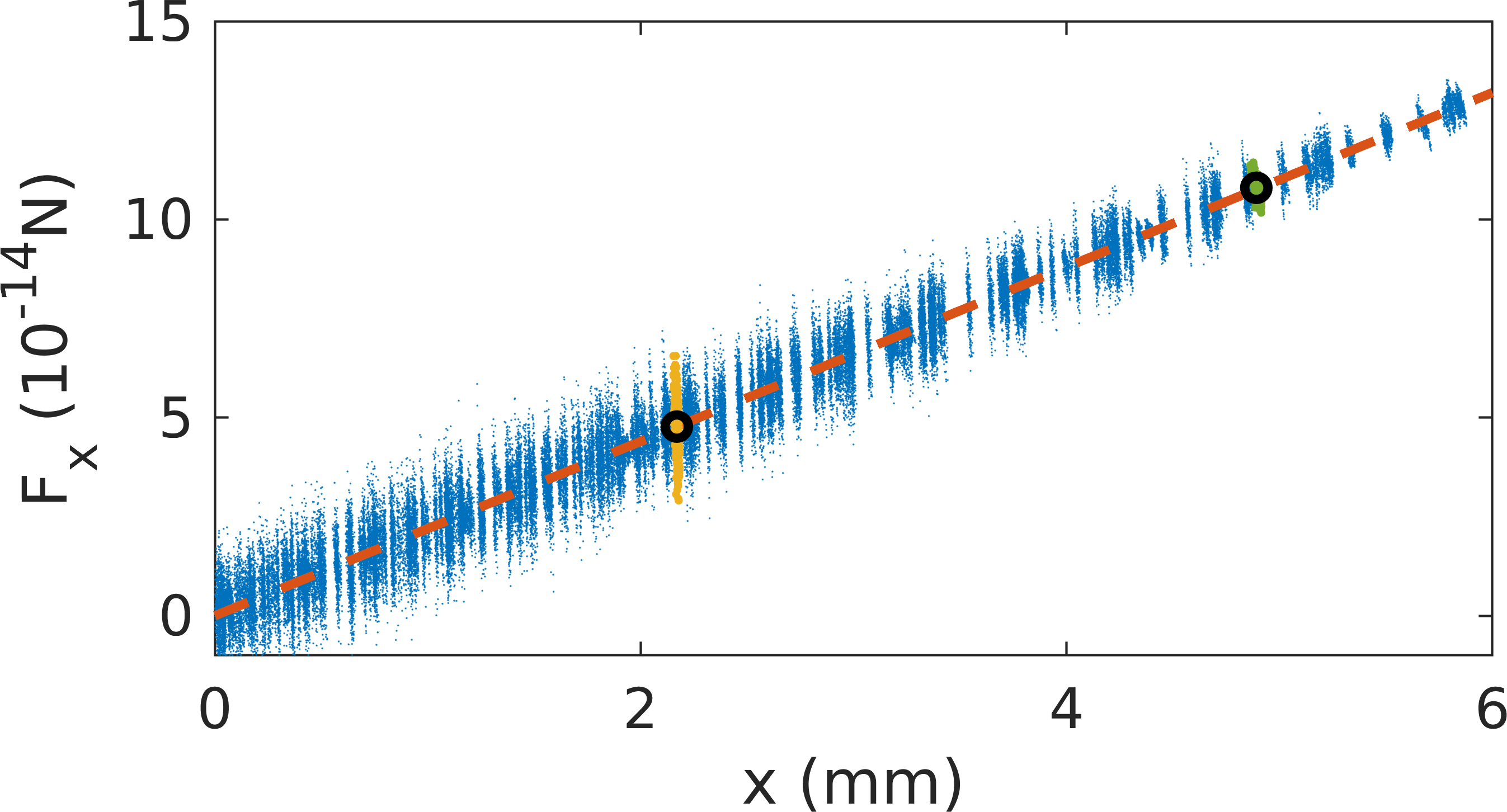}
\caption{Residual $x$ component of the force $F_x$ of particles as a function of distance from the cluster center in $x$ direction in a simulated dust cluster. The mean position of two particles are indicated by black circles and their respective forces are highlighted. It can be seen that the mean positions, where the particles are supposed to be in an equilibrium state, can be modeled well by a linear fit.}
\label{fig_check1}
\end{figure}
\begin{figure}
\includegraphics[width=\columnwidth]{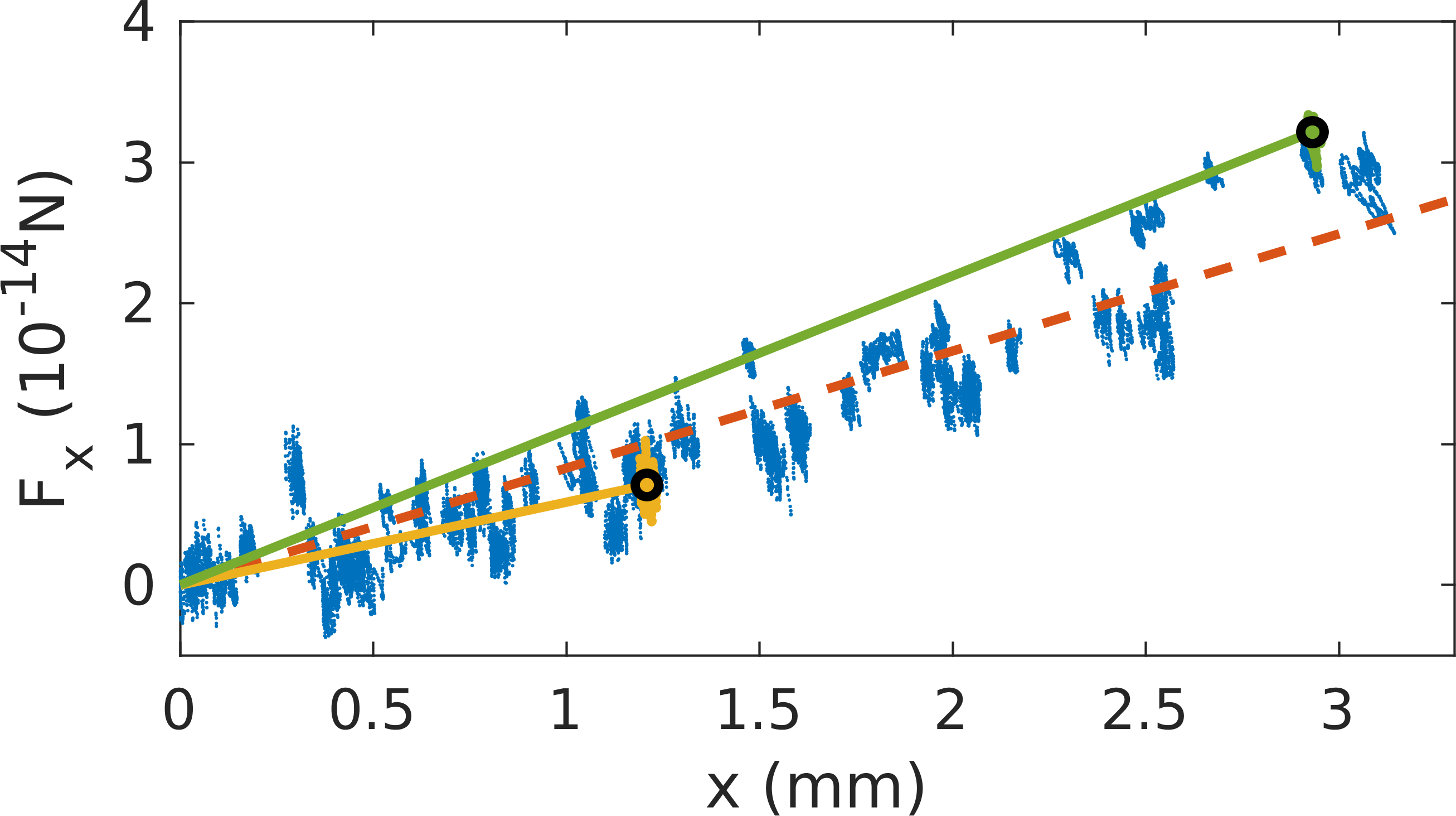}
\caption{Residual $x$ component of the force $F_x$ of particles  as a function of distance from the cluster center in $x$ direction in a measured dust cluster. The mean position of two particles are indicated by black circles and their respective forces are highlighted. The mean positions of the particles can not be modeled by a single linear fit (dashed line). Instead, a linear fit is made with every particle's mean position which is indicated for two particles by the two solid lines.}
\label{fig_check2}
\end{figure}
Before experimental measurements are investigated, it is necessary to check the model for the confinement force. Its validity is crucial for quantitatively accurate results. The basic idea behind the configurational temperature is, that the particles are in equilibrium, hence, the mean force on each particle should vanish. The configurational temperature then follows from the deviations of this equilibrium position against the interparticle as well as the confinement forces. 

Figure\,\ref{fig_check1} shows, again for the MD simulation, the particle interaction forces on every particle $\mathbf{F}_j^\mathrm{el}$ according to Eq.\,(\ref{eq_forceSum}). They are computed using the simulation parameters for the charge and screening length. One can see, that a confinement force, linearly increasing from the center, is necessary to result in a thermal fluctuation around the force equilibrium. Such  a linear confinement force is expected for a harmonic confinement. The slope of the force is the same for all particles and agrees with the value of $k$ chosen in the simulation. It is interesting to note here that the random thermal excursions are much larger near the cluster center, since there the total force is smallest.

However, the situation is different when a laboratory dust cluster is investigated. Figure\,\ref{fig_check2} shows the particle interaction force in a laboratory cluster. Here the confinement force on the dust particles is not very well modelled by a global harmonic confinement (dashed line). Instead, it is more reasonable to assign a local harmonic confinement. There, an individual confinement force constant $k_j$ is assigned to each particles  as indicated by the solid lines for two exemplary particles. Since the configurational temperature is determined from the random excursions from the equilibrium, using the global harmonic confinement would result in a large overestimation of the temperature. Using the local slope, only the excursions from the local equilibrium enter the configurational temperature. With this modification, Eq.\,(\ref{eq_CTfinal}) can be written as
\begin{equation}\label{eq_CTind}
k_\mathrm{B} T_\mathrm{conf} = -\frac{\left\langle \sum_{j=1}^N  \left(\mathbf{F}_j^\mathrm{el} - \mathbf{q}_j k_j \right)^2 \right\rangle}{\left\langle\sum_{j=1}^N \left(  \nabla_j \cdot \mathbf{F}_j^\mathrm{el}  + D k_j \right)   \right\rangle}.
\end{equation}
For two exemplary particles from Fig.\,\ref{fig_check2}, this individual confinement force is indicated by solid lines.

\subsection{Laboratory measurements}
\begin{figure}
\includegraphics[width=\columnwidth]{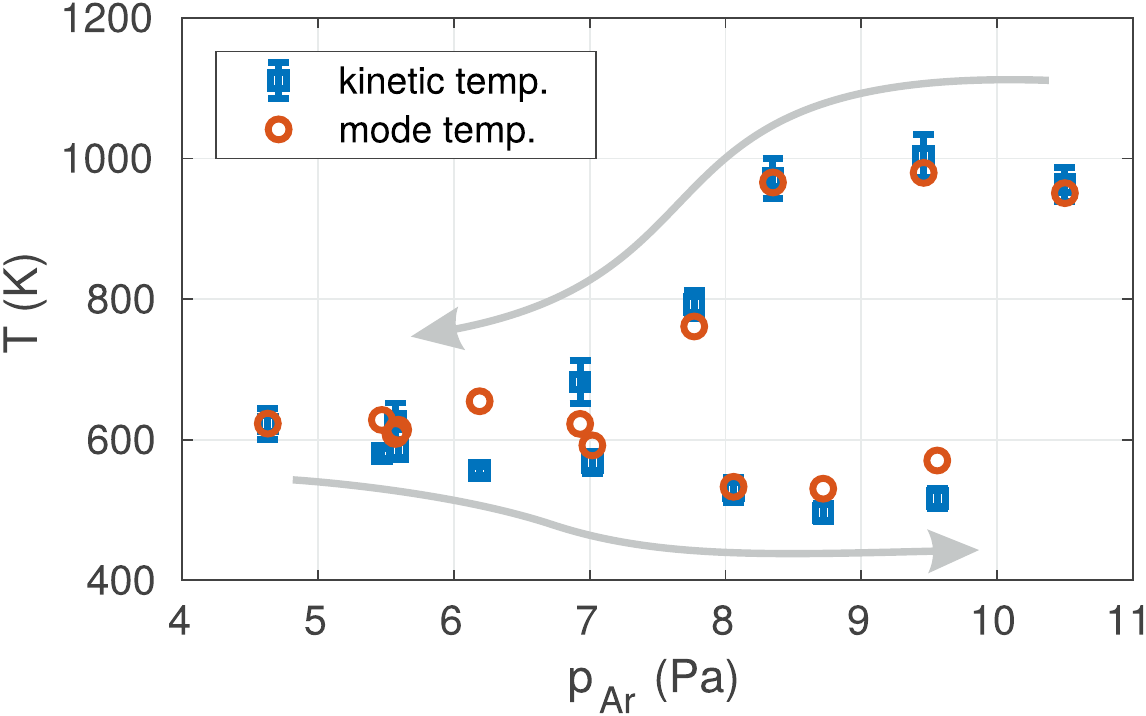}
\caption{The cluster temperature is shown depending on the neutral gas pressure. The blue squares indicate the kinetic temperature with error bars, the yellow circles represent the mode temperature values.}
\label{fig_tempVSpres}
\end{figure}
\begin{figure}
\includegraphics[width=\columnwidth]{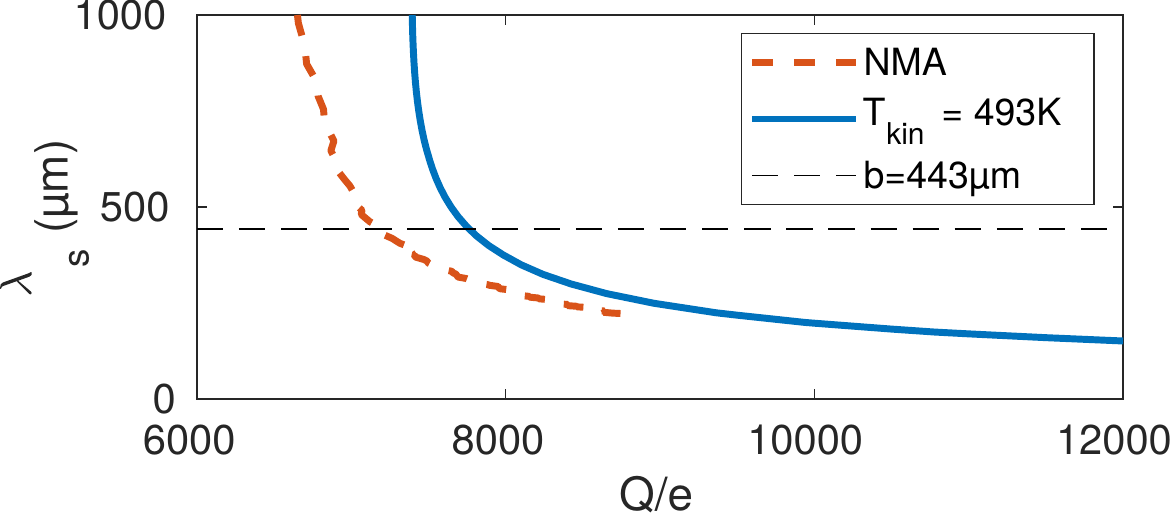}
\caption{$\lambda_\mathrm{s}(Q)$-functions for a measured dust cluster at $p=8.2\,\mathrm{Pa}$. The solid curve is retrieved by intersecting the configurational temperature surface at the measured kinetic temperature of $493\,\mathrm{K}$. The thick dashed line is obtained from the mode analysis of the cluster. The thin dashed line indicates the measured mean interparticle distance.}
\label{fig_mess9}
\end{figure}
The laboratory measurements have been done with a 155 particle cluster under varied neutral gas pressure. In our case, the recorded sequences have the quality to determine also the particle velocities and hence their kinetic temperature. In Fig.\,\ref{fig_tempVSpres}, the kinetic temperature measured in the cluster is shown together with the mode temperature determined from NMA. The determination of the cluster temperature is quite consistent using either the mode temperature or the kinetic temperature (which is no real surprise since both measurements rely on the particle velocities). In the experiment, the gas pressure has been first reduced from 11 to $5\,\mathrm{Pa}$ resulting in a slight decrease of dust temperature. When the gas pressure is increased again, a further slight reduction in temperature is seen. This behavior is due to the fact, that our discharge operates in two slightly different discharge modes. When changing the gas pressure the discharge changes between the two modes with hysteresis. This allows us to realize dust clusters at different kinetic temperatures which then can be analyzed using the configurational temperature approach.  

Taking the specific example of the cluster at $8.2\,\mathrm{Pa}$, the kinetic temperature was found to be $493\,\mathrm{K}$. As the kinetic temperatures is known, it is also possible to compare it with the configurational temperatures. Here, the configurational temperature is calculated on a relevant $\lambda_\mathrm{s}$-$Q$-grid. Further in the calculation of the configurational temperature according to Eq.\,(\ref{eq_CTind}), the local confinement strengths $k_j$ are used. From the comparison with the kinetic temperature the matching $\lambda_\mathrm{s}-Q$-pairs are derived similar to Fig.\,\ref{fig_surf}. For this measurement, Fig.\,\ref{fig_mess9} shows the corresponding curve together with the $\lambda_\mathrm{s}(Q)$ obtained by NMA. In contrast to the simulated cluster (see Fig.\,\ref{fig_methods}), the two curves somewhat deviate from each other. It can be seen that the configurational temperature approach suggests somewhat higher particle charges. The difference to NMA might lie in the fact, that NMA assumes a harmonic confinement in Eqs.\,(\ref{nma1})--(\ref{nma3}). Approximating again the screening length by the interparticle distance, i.e., $\lambda_\mathrm{s} \approx b=443\upmu\mathrm{m}$, the particle charge can be estimated as $7200e$ for the NMA analysis and $7900e$ for the configurational temperature analysis. 

\begin{figure}
\includegraphics[width=\columnwidth]{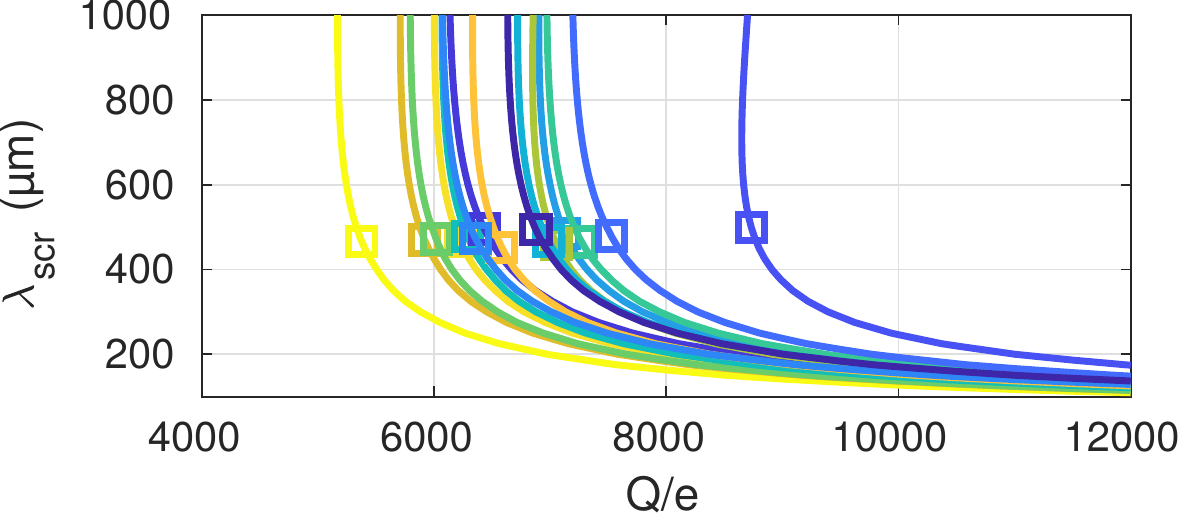}
\caption{The different $\lambda_\mathrm{s}(Q)$-curves are obtained from a dust cluster under different neutral gas pressures. The pressure is color-coded and increases from blue to yellow color just as in Fig.\,\ref{fig_qpres}. The function values using the nearest-neighbor distances between the particles as a measure for $\lambda_\mathrm{s}$ are highlighted by squares.}
\label{fig_lQ}
\end{figure}

\begin{figure}
\includegraphics[width=\columnwidth]{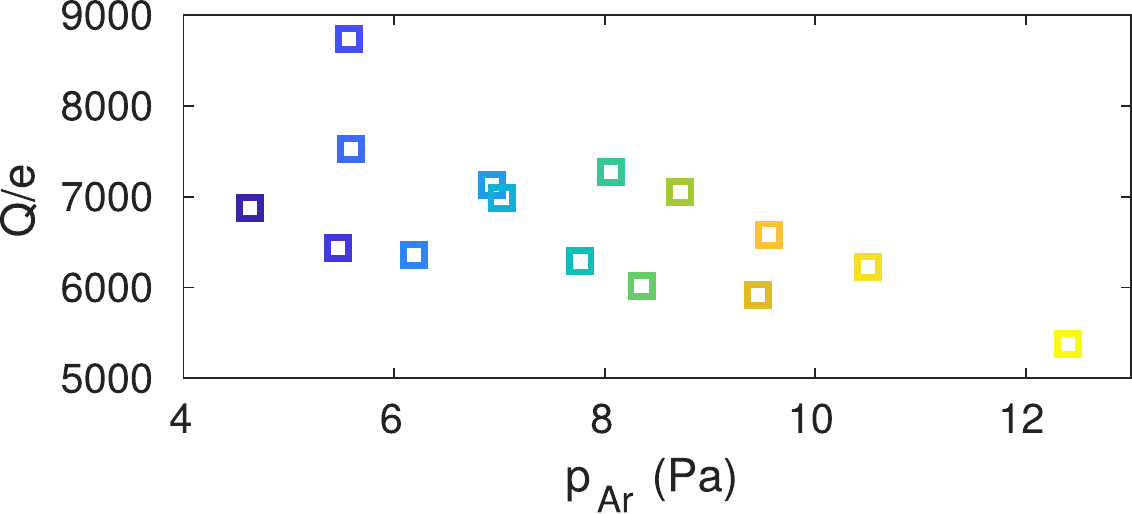}
\caption{The particle charge dependence on the neutral gas pressure. The color of the markers reflect the pressure accordingly.}
\label{fig_qpres}
\end{figure}

For further analysis of the measurements, the configurational temperature approach is used for the measured clusters at all pressures. By taking the kinetic temperatures from Fig.\,\ref{fig_tempVSpres} for each   measurement, the $\lambda_\mathrm{s}(Q)$-lines for each cluster have been extracted from the configurational temperature surfaces. These lines are shown in Fig.\,\ref{fig_lQ}. 
The lines look very similar to each other, with mainly their horizontal position being shifted. Now, the particle charges in the different situations have been extracted by assuming that the mean interparticle distance $b$ reflects the screening length $\lambda_\mathrm{s}$, i.e., as above we assume $\lambda_\mathrm{s}\approx b$. The corresponding points on the graphs are indicated by squares. Note that even deviations of $100\,\mathrm{\upmu m}$ in $\lambda_\mathrm{s}$ (or $b$) only have a slight influence of about $\Delta Q = \pm 100e$ in the determined charge. The extracted particle charges are shown in Fig.\,\ref{fig_qpres} as a function of the neutral gas pressure of the corresponding measurement. It is seen that the particles get less negatively charged with increasing neutral gas pressure, which is expected from an increased collisional ion current at higher pressures\,\cite{khrapak}.

\subsection{Three-dimensional systems}
In the previous sections, the systems that have been analyzed were two-dimensional. When all two-dimensional interactions and confinement forces acting on the particles are known, the position fluctuations can be used to calculate the configurational temperature. When the dust cluster has a three-dimensional structure instead, the knowledge of the three-dimensional interaction and confinement forces is necessary. Thus, it is clear that imaging diagnostics based on a two-dimensional section of a large dust cloud can not be used for the configurational temperature approach. Considering the state-of-the-art of modern imaging equipment, it is not reasonable to assume that all particle trajectories in a large 3D dust cloud can be reliably determined. Thus, we will show how to apply the configurational temperature approach in large dust systems without the need for determining all particle positions but only a subset.

Despite the fact that we have and need the 3D positions, we will evaluate the configurational temperature from a 2D projection of the forces. This is mainly due to the fact that the illuminated volume is relatively thin in $z$-direction. When in Eq.\,(\ref{eq_CTfinal}) only two components of the three-dimensional system are considered, the resulting two-dimensional configurational temperature
\begin{equation}\label{eq_confT2d}
 -\frac{\left\langle \sum_{j=1}^N  \left(  F_{j,x} +F_{j, y} \right)^2 \right\rangle}{\left\langle\sum_{j=1}^N  \left( \partial_\mathrm{j, x} F_{j,x} + \partial_{j,y} F_{j,y} \right) \right\rangle} = k_\mathrm{B} T_\mathrm{conf}
\end{equation}
does also return the actual configurational temperature of the system. It should be noted, that for calculating the forces $F_x$ and $F_y$, the full 3D positions are required. Further, in a trade-off with statistical accuracy, the sum can be computed over a subset of particles, only. And last, the forces and their divergence decrease exponentially with distance, which suggests that particles that are far away from the considered ensemble do not influence the calculation. 

\begin{figure}
\includegraphics[width=\columnwidth]{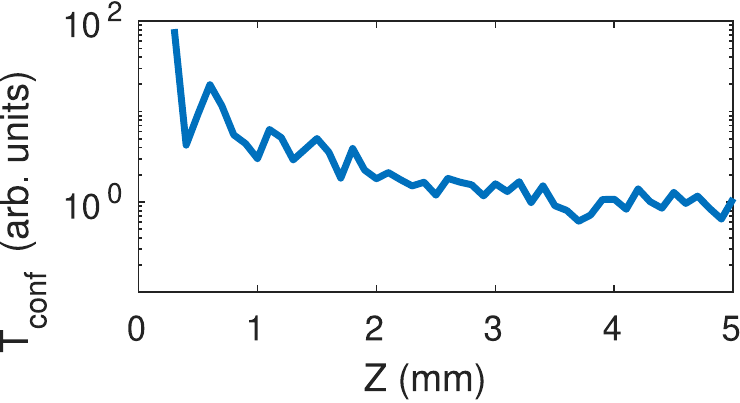}
\caption{Configurational temperature of central particles in a cuboid volume, depending on the total thickness of this volume.}
\label{fig_thickness}
\end{figure}

To compute reliable configurational temperatures in 3D, we first have checked what distance in $z$-direction has to be included in the calculation. Due to shielding, neglecting particles far away from a considered slice do not significantly influence the forces. Such a test is shown in Fig.\,\ref{fig_thickness}.
For this benchmark test, simulated particle trajectories have been used. The particles are initially placed on a three-dimensional face-centered-cubic grid with interparticle spacing of $500\,\upmu \mathrm{m}$. Then, the trajectories are evolved from these starting positions by adding Gaussian distributed noise to the particle positions. As a result, the particles seem to reflect a Brownian motion with a constant temperature. From this set, particles that lie inside a cuboid with dimensions $10\times 10 \times Z\,\mathrm{mm}^3$ are selected, with $Z$ being the thickness of this volume. In the calculation of the forces, all particles from the entire volume of thickness $Z$ are taken into account. The configurational temperature is however calculated only for particles in a thin central slice with $z= 0 \pm 0.1\,\mathrm{mm}$. 

It can be seen, that the derived configurational temperature is overestimated when only particles within a very thin volume are considered for force contribution. This is since the average force equilibrium position does not coincide with the average particle position due exclusion of particles with a non negligible influence. From approximately $Z > 2\,\mathrm{mm}$ on, the considered volume for force contribution is sufficiently large to account for all relevant particle interactions. Hence, a constant temperature level is reached. Adding the interactions from particles at distances to the central plane larger than $\pm 1\,\mathrm{mm}$ (2 interparticle distances), does not further improve the derived configurational temperature. Thus, if a volumetric particle ensemble is investigated, particle positions from a volume with a thickness of at least 4 interparticle distances have to be known for the range of parameters $Q$ and $\lambda_\mathrm{s}$ considered here. It should be mentioned again that the approach depends on the knowledge of the three-dimensional interaction forces between the particles, but using Eq.\,(\ref{eq_confT2d}) the configurational temperature is calculated from the 2D projections of these 3D forces.

An experimental realization of this approach is presented in the following.

\subsection{Three-dimensional measurements under microgravity}
\begin{figure}
\includegraphics[width=\columnwidth]{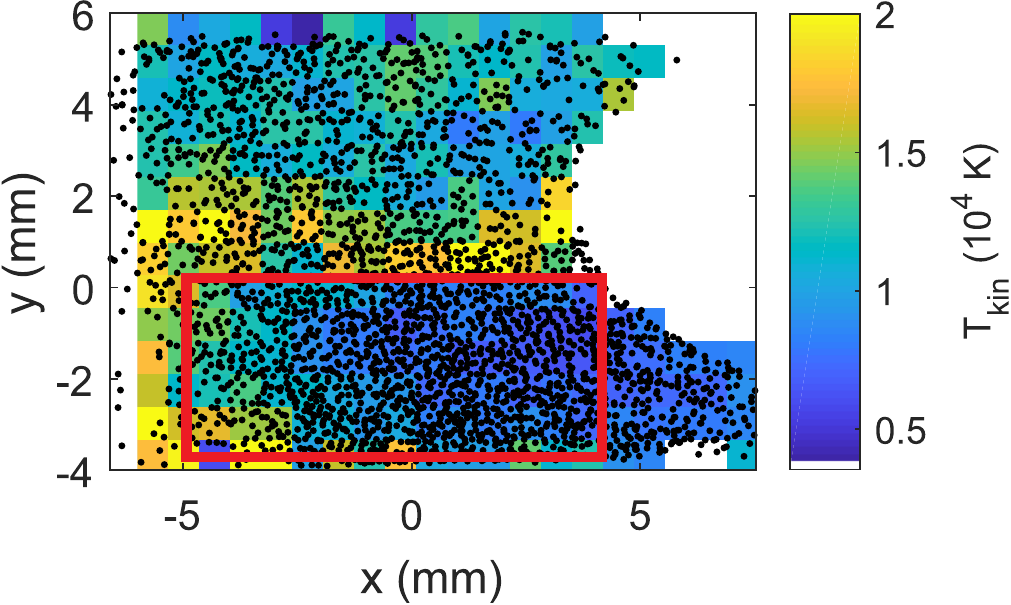}
\caption{Kinetic temperature in a section through a large dust cloud. The $x$- and $y$-positions of about 3000 detected particles from a single frame are shown as black dots. On the right side one can identify the particle free region in the center of the discharge (compare Fig.\,\ref{fig_experiment}).}
\label{fig_kinmap}
\end{figure}
For this investigation, a dust cloud is formed within a symmetric discharge as shown in Fig.\,\ref{fig_experiment}(b). The three-dimensional position measurements are made using an illumination laser sheet with a thickness of about $2\,\mathrm{mm}$. As the interparticle distance is about $350\,\upmu \mathrm{m}$, this complies with the   condition that the thickness of the illuminated volume should exceed four times the interparticle distance derived in the previous section. Thus, the three-dimensional particle positions can be used to determine the  configurational temperature properly. In the $x$ and $y$-direction, the field of view is $14\times 10\,\mathrm{mm}$. As the measurements are made at a framerate of $200\,\mathrm{fps}$, the kinetic temperature can also be determined. To get an insight into the temperature distribution in this data, the measured particle velocities are gathered in discretized regions of the measurement volume. For each region, the kinetic temperature is calculated. The result is shown in Fig.\,\ref{fig_kinmap}. Regions (such as the void) with no or very few detected particles are left white.

For further analysis we restrict ourselves to the data in a region with a constant temperature (red rectangle in Fig.\,\ref{fig_kinmap}). The size of this volume is $9\times 4 \times 2\,\mathrm{mm}^3$. Particles from within this volume are used for force contributions. The inner particles which are considered for computing the configurational temperature are from a $2\times 0.5 \times 0.25\,\mathrm{mm}^3$-region centered in the investigated volume. Further, to guarantee a reliable force equilibrium, only particles in the inner slice that have at least 12 neighbors within a distance of $400\,\mathrm{\upmu m}$ have been included in the computation of the configurational temperature (12 is the number of nearest neighbors in a densely packed system). 

\begin{figure}
\includegraphics[width=\columnwidth]{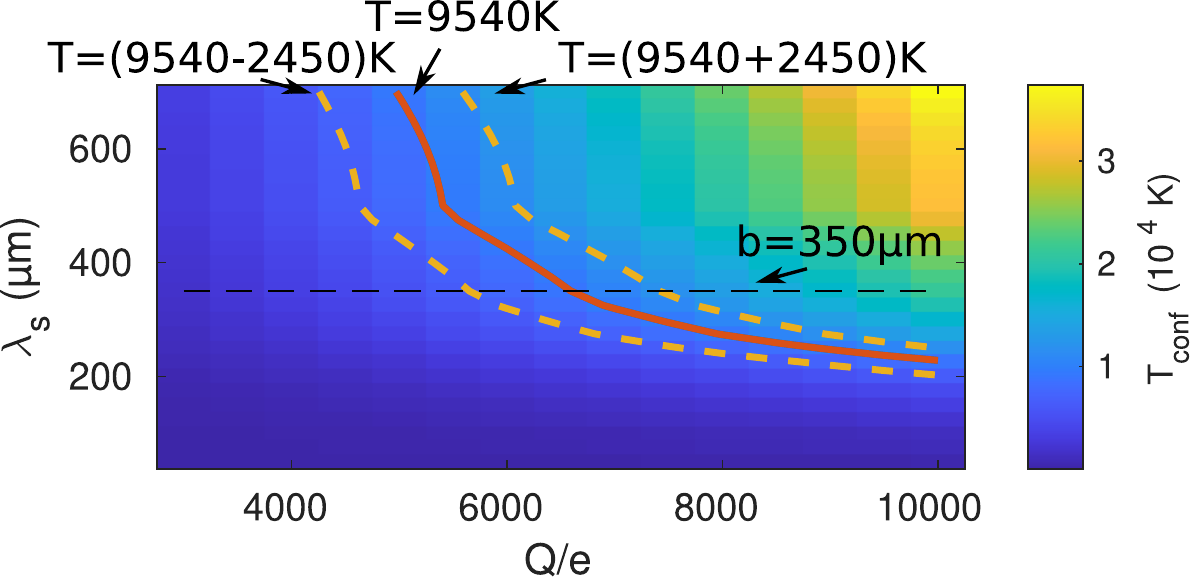}
\caption{Configurational temperature of a volumetric measurement in a dust cloud. The solid line indicates the $\lambda_\mathrm{s}(Q)$ function at a given temperature of $9540 \pm 2450\,\mathrm{K}$.}
\label{fig_pfcmap}
\end{figure}

The mean kinetic temperature for the observation volume was determined to be $T_\mathrm{kin} = 9540 \pm 2450\,\mathrm{K}$. Figure\,\ref{fig_pfcmap} again shows the lines $\lambda_\mathrm{s}(Q)$ where the configurational temperature coincides with the measured kinetic temperature (including error range). The central solid line indicates the mean temperature and the dashed lines show the confidence interval. The particle charge is found to be about $Q \approx 6500e \pm 1000e$ at a screening length $\lambda_\mathrm{s}$ that we again have estimated to be close to the interparticle distance $b = 350\,\upmu\mathrm{m}$. This obtained charge does not considerably change even when taking $\lambda_\mathrm{s} = 2b = 700\upmu\mathrm{m}$.

To obtain a theoretical estimate of the particle charge, we compare with self-consistent computations that have been done for a very similar gas discharge\,\cite{akdim_plasma}. There, the electrode voltages were slightly lower than ours ($70\,\mathrm{V}$ compared to $76\,\mathrm{V}$) and the pressure was slightly higher than our ($40\,\mathrm{Pa}$ compared to $30\,\mathrm{Pa}$). The plasma parameters that we employ from these calculations are the electron density $n_{e0} = 2.2\cdot 10^{14}$ and the electron temperature $T_e = 4\,\mathrm{eV}$. The charge then can be computed with respect to electron depletion effects\,\cite{havnes} and with a fixed ion density\,\cite{Goertz11} $n_i = n_{e,0}$ to be $Q=9800e$. Regarding the slightly different discharge geometry and operation parameters, the computed particle charge compares quite well to our measured particle charge in the dust cloud under microgravity.

\section{Summary and Conclusions}
In this paper, we have introduced the configurational temperature as a tool to analyze dusty plasma systems. The measurement of the configurational temperature is based on the determination of the particle positions, rather than the particle velocities which may be favourable for some experimental situations. The concept of configurational temperature was discussed and applied to simulated and experimental data. There, it has been found that the proper modeling of the particle confinement is an important step to consider and that the approach is consistent with established methods.

Furthermore, the configurational temperature in combination with the kinetic temperature has been used to determine the particle charge. Such non-invasive measurements of the particle charge are of great interest in the field of dusty plasmas. The proposed algorithm was demonstrated to work for two-dimensional and three-dimensional measurements. 

\section{Acknowledgements}
Financial support from the German Aerospace Center (DLR) under Project No. 50WM1638 is gratefully acknowledged.
%


\begin{thebibliography}{47}%
\makeatletter
\providecommand \@ifxundefined [1]{%
 \@ifx{#1\undefined}
}%
\providecommand \@ifnum [1]{%
 \ifnum #1\expandafter \@firstoftwo
 \else \expandafter \@secondoftwo
 \fi
}%
\providecommand \@ifx [1]{%
 \ifx #1\expandafter \@firstoftwo
 \else \expandafter \@secondoftwo
 \fi
}%
\providecommand \natexlab [1]{#1}%
\providecommand \enquote  [1]{``#1''}%
\providecommand \bibnamefont  [1]{#1}%
\providecommand \bibfnamefont [1]{#1}%
\providecommand \citenamefont [1]{#1}%
\providecommand \href@noop [0]{\@secondoftwo}%
\providecommand \href [0]{\begingroup \@sanitize@url \@href}%
\providecommand \@href[1]{\@@startlink{#1}\@@href}%
\providecommand \@@href[1]{\endgroup#1\@@endlink}%
\providecommand \@sanitize@url [0]{\catcode `\\12\catcode `\$12\catcode
  `\&12\catcode `\#12\catcode `\^12\catcode `\_12\catcode `\%12\relax}%
\providecommand \@@startlink[1]{}%
\providecommand \@@endlink[0]{}%
\providecommand \url  [0]{\begingroup\@sanitize@url \@url }%
\providecommand \@url [1]{\endgroup\@href {#1}{\urlprefix }}%
\providecommand \urlprefix  [0]{URL }%
\providecommand \Eprint [0]{\href }%
\providecommand \doibase [0]{http://dx.doi.org/}%
\providecommand \selectlanguage [0]{\@gobble}%
\providecommand \bibinfo  [0]{\@secondoftwo}%
\providecommand \bibfield  [0]{\@secondoftwo}%
\providecommand \translation [1]{[#1]}%
\providecommand \BibitemOpen [0]{}%
\providecommand \bibitemStop [0]{}%
\providecommand \bibitemNoStop [0]{.\EOS\space}%
\providecommand \EOS [0]{\spacefactor3000\relax}%
\providecommand \BibitemShut  [1]{\csname bibitem#1\endcsname}%
\let\auto@bib@innerbib\@empty
\bibitem [{\citenamefont {Quinn}\ and\ \citenamefont {Goree}(2000)}]{QuinnPRE}%
  \BibitemOpen
  \bibfield  {author} {\bibinfo {author} {\bibfnamefont {R.~A.}\ \bibnamefont
  {Quinn}}\ and\ \bibinfo {author} {\bibfnamefont {J.}~\bibnamefont {Goree}},\
  }\href {\doibase 10.1103/PhysRevE.61.3033} {\bibfield  {journal} {\bibinfo
  {journal} {Phys. Rev. E}\ }\textbf {\bibinfo {volume} {61}},\ \bibinfo
  {pages} {3033} (\bibinfo {year} {2000})}\BibitemShut {NoStop}%
\bibitem [{\citenamefont {Nunomura}\ \emph
  {et~al.}(2002{\natexlab{a}})\citenamefont {Nunomura}, \citenamefont {Goree},
  \citenamefont {Hu}, \citenamefont {Wang}, \citenamefont {Bhattacharjee},\
  and\ \citenamefont {Avinash}}]{NunomuraPRL}%
  \BibitemOpen
  \bibfield  {author} {\bibinfo {author} {\bibfnamefont {S.}~\bibnamefont
  {Nunomura}}, \bibinfo {author} {\bibfnamefont {J.}~\bibnamefont {Goree}},
  \bibinfo {author} {\bibfnamefont {S.}~\bibnamefont {Hu}}, \bibinfo {author}
  {\bibfnamefont {X.}~\bibnamefont {Wang}}, \bibinfo {author} {\bibfnamefont
  {A.}~\bibnamefont {Bhattacharjee}}, \ and\ \bibinfo {author} {\bibfnamefont
  {K.}~\bibnamefont {Avinash}},\ }\href {\doibase
  10.1103/PhysRevLett.89.035001} {\bibfield  {journal} {\bibinfo  {journal}
  {Phys. Rev. Lett.}\ }\textbf {\bibinfo {volume} {89}},\ \bibinfo {pages}
  {035001} (\bibinfo {year} {2002}{\natexlab{a}})}\BibitemShut {NoStop}%
\bibitem [{\citenamefont {Schmidt}\ and\ \citenamefont
  {Piel}(2015)}]{schmidt_heating}%
  \BibitemOpen
  \bibfield  {author} {\bibinfo {author} {\bibfnamefont {C.}~\bibnamefont
  {Schmidt}}\ and\ \bibinfo {author} {\bibfnamefont {A.}~\bibnamefont {Piel}},\
  }\href {\doibase 10.1103/PhysRevE.92.043106} {\bibfield  {journal} {\bibinfo
  {journal} {Phys. Rev. E}\ }\textbf {\bibinfo {volume} {92}},\ \bibinfo
  {pages} {043106} (\bibinfo {year} {2015})}\BibitemShut {NoStop}%
\bibitem [{\citenamefont {Fortov}\ \emph {et~al.}(2007)\citenamefont {Fortov},
  \citenamefont {Vaulina}, \citenamefont {Petrov}, \citenamefont {Vasiliev},
  \citenamefont {Gavrikov}, \citenamefont {Shakova}, \citenamefont {Vorona},
  \citenamefont {Khrustalyov}, \citenamefont {Manohin},\ and\ \citenamefont
  {Chernyshev}}]{heat}%
  \BibitemOpen
  \bibfield  {author} {\bibinfo {author} {\bibfnamefont {V.~E.}\ \bibnamefont
  {Fortov}}, \bibinfo {author} {\bibfnamefont {O.~S.}\ \bibnamefont {Vaulina}},
  \bibinfo {author} {\bibfnamefont {O.~F.}\ \bibnamefont {Petrov}}, \bibinfo
  {author} {\bibfnamefont {M.~N.}\ \bibnamefont {Vasiliev}}, \bibinfo {author}
  {\bibfnamefont {A.~V.}\ \bibnamefont {Gavrikov}}, \bibinfo {author}
  {\bibfnamefont {I.~A.}\ \bibnamefont {Shakova}}, \bibinfo {author}
  {\bibfnamefont {N.~A.}\ \bibnamefont {Vorona}}, \bibinfo {author}
  {\bibfnamefont {Y.~V.}\ \bibnamefont {Khrustalyov}}, \bibinfo {author}
  {\bibfnamefont {A.~A.}\ \bibnamefont {Manohin}}, \ and\ \bibinfo {author}
  {\bibfnamefont {A.~V.}\ \bibnamefont {Chernyshev}},\ }\href {\doibase
  10.1103/PhysRevE.75.026403} {\bibfield  {journal} {\bibinfo  {journal} {Phys.
  Rev. E}\ }\textbf {\bibinfo {volume} {75}},\ \bibinfo {pages} {026403}
  (\bibinfo {year} {2007})}\BibitemShut {NoStop}%
\bibitem [{\citenamefont {Nosenko}\ \emph {et~al.}(2006)\citenamefont
  {Nosenko}, \citenamefont {Goree},\ and\ \citenamefont {Piel}}]{NosenkoPOP13}%
  \BibitemOpen
  \bibfield  {author} {\bibinfo {author} {\bibfnamefont {V.}~\bibnamefont
  {Nosenko}}, \bibinfo {author} {\bibfnamefont {J.}~\bibnamefont {Goree}}, \
  and\ \bibinfo {author} {\bibfnamefont {A.}~\bibnamefont {Piel}},\ }\href
  {\doibase 10.1063/1.2182207} {\bibfield  {journal} {\bibinfo  {journal}
  {Physics of Plasmas}\ }\textbf {\bibinfo {volume} {13}},\ \bibinfo {pages}
  {032106} (\bibinfo {year} {2006})},\ \Eprint
  {http://arxiv.org/abs/https://doi.org/10.1063/1.2182207}
  {https://doi.org/10.1063/1.2182207} \BibitemShut {NoStop}%
\bibitem [{\citenamefont {Schablinski}\ \emph {et~al.}(2012)\citenamefont
  {Schablinski}, \citenamefont {Block}, \citenamefont {Piel}, \citenamefont
  {Melzer}, \citenamefont {Thomsen}, \citenamefont {K{\"a}hlert},\ and\
  \citenamefont {Bonitz}}]{SchablinskiPOP19}%
  \BibitemOpen
  \bibfield  {author} {\bibinfo {author} {\bibfnamefont {J.}~\bibnamefont
  {Schablinski}}, \bibinfo {author} {\bibfnamefont {D.}~\bibnamefont {Block}},
  \bibinfo {author} {\bibfnamefont {A.}~\bibnamefont {Piel}}, \bibinfo {author}
  {\bibfnamefont {A.}~\bibnamefont {Melzer}}, \bibinfo {author} {\bibfnamefont
  {H.}~\bibnamefont {Thomsen}}, \bibinfo {author} {\bibfnamefont
  {H.}~\bibnamefont {K{\"a}hlert}}, \ and\ \bibinfo {author} {\bibfnamefont
  {M.}~\bibnamefont {Bonitz}},\ }\href {\doibase 10.1063/1.3677356} {\bibfield
  {journal} {\bibinfo  {journal} {Physics of Plasmas}\ }\textbf {\bibinfo
  {volume} {19}},\ \bibinfo {pages} {013705} (\bibinfo {year} {2012})},\
  \Eprint {http://arxiv.org/abs/https://doi.org/10.1063/1.3677356}
  {https://doi.org/10.1063/1.3677356} \BibitemShut {NoStop}%
\bibitem [{\citenamefont {Hubertus M.~Thomas}(1996)}]{ThomasNature}%
  \BibitemOpen
  \bibfield  {author} {\bibinfo {author} {\bibnamefont {H. M.~Thomas}\
  \bibfnamefont {G.~E.~Morfill} },\ }\href@noop {} {\bibfield  {journal}
  {\bibinfo  {journal} {Nature}\ }\textbf {\bibinfo {volume} {379}},\ \bibinfo
  {pages} {806} (\bibinfo {year} {1996})}\BibitemShut {NoStop}%
\bibitem [{\citenamefont {Melzer}\ \emph {et~al.}(1996)\citenamefont {Melzer},
  \citenamefont {Homann},\ and\ \citenamefont {Piel}}]{MelzerPRE53}%
  \BibitemOpen
  \bibfield  {author} {\bibinfo {author} {\bibfnamefont {A.}~\bibnamefont
  {Melzer}}, \bibinfo {author} {\bibfnamefont {A.}~\bibnamefont {Homann}}, \
  and\ \bibinfo {author} {\bibfnamefont {A.}~\bibnamefont {Piel}},\ }\href
  {\doibase 10.1103/PhysRevE.53.2757} {\bibfield  {journal} {\bibinfo
  {journal} {Phys. Rev. E}\ }\textbf {\bibinfo {volume} {53}},\ \bibinfo
  {pages} {2757} (\bibinfo {year} {1996})}\BibitemShut {NoStop}%
\bibitem [{\citenamefont {Hayashi}\ and\ \citenamefont
  {Tachibana}(1994)}]{Hayashi_1994}%
  \BibitemOpen
  \bibfield  {author} {\bibinfo {author} {\bibfnamefont {Y.}~\bibnamefont
  {Hayashi}}\ and\ \bibinfo {author} {\bibfnamefont {K.}~\bibnamefont
  {Tachibana}},\ }\href {\doibase 10.1143/jjap.33.l804} {\bibfield  {journal}
  {\bibinfo  {journal} {Japanese Journal of Applied Physics}\ }\textbf
  {\bibinfo {volume} {33}},\ \bibinfo {pages} {L804} (\bibinfo {year}
  {1994})}\BibitemShut {NoStop}%
\bibitem [{\citenamefont {Chu}\ and\ \citenamefont {I}(1994)}]{ChuPRL72}%
  \BibitemOpen
  \bibfield  {author} {\bibinfo {author} {\bibfnamefont {J.~H.}\ \bibnamefont
  {Chu}}\ and\ \bibinfo {author} {\bibfnamefont {L.}~\bibnamefont {I}},\ }\href
  {\doibase 10.1103/PhysRevLett.72.4009} {\bibfield  {journal} {\bibinfo
  {journal} {Phys. Rev. Lett.}\ }\textbf {\bibinfo {volume} {72}},\ \bibinfo
  {pages} {4009} (\bibinfo {year} {1994})}\BibitemShut {NoStop}%
\bibitem [{\citenamefont {Thomas}\ \emph {et~al.}(1994)\citenamefont {Thomas},
  \citenamefont {Morfill}, \citenamefont {Demmel}, \citenamefont {Goree},
  \citenamefont {Feuerbacher},\ and\ \citenamefont {M\"ohlmann}}]{crystall}%
  \BibitemOpen
  \bibfield  {author} {\bibinfo {author} {\bibfnamefont {H.}~\bibnamefont
  {Thomas}}, \bibinfo {author} {\bibfnamefont {G.~E.}\ \bibnamefont {Morfill}},
  \bibinfo {author} {\bibfnamefont {V.}~\bibnamefont {Demmel}}, \bibinfo
  {author} {\bibfnamefont {J.}~\bibnamefont {Goree}}, \bibinfo {author}
  {\bibfnamefont {B.}~\bibnamefont {Feuerbacher}}, \ and\ \bibinfo {author}
  {\bibfnamefont {D.}~\bibnamefont {M\"ohlmann}},\ }\href {\doibase
  10.1103/PhysRevLett.73.652} {\bibfield  {journal} {\bibinfo  {journal} {Phys.
  Rev. Lett.}\ }\textbf {\bibinfo {volume} {73}},\ \bibinfo {pages} {652}
  (\bibinfo {year} {1994})}\BibitemShut {NoStop}%
\bibitem [{\citenamefont {Dietz}\ \emph {et~al.}(2018)\citenamefont {Dietz},
  \citenamefont {Bergert}, \citenamefont {Steinm\"uller}, \citenamefont
  {Kretschmer}, \citenamefont {Mitic},\ and\ \citenamefont {Thoma}}]{DietzPRE}%
  \BibitemOpen
  \bibfield  {author} {\bibinfo {author} {\bibfnamefont {C.}~\bibnamefont
  {Dietz}}, \bibinfo {author} {\bibfnamefont {R.}~\bibnamefont {Bergert}},
  \bibinfo {author} {\bibfnamefont {B.}~\bibnamefont {Steinm\"uller}}, \bibinfo
  {author} {\bibfnamefont {M.}~\bibnamefont {Kretschmer}}, \bibinfo {author}
  {\bibfnamefont {S.}~\bibnamefont {Mitic}}, \ and\ \bibinfo {author}
  {\bibfnamefont {M.~H.}\ \bibnamefont {Thoma}},\ }\href {\doibase
  10.1103/PhysRevE.97.043203} {\bibfield  {journal} {\bibinfo  {journal} {Phys.
  Rev. E}\ }\textbf {\bibinfo {volume} {97}},\ \bibinfo {pages} {043203}
  (\bibinfo {year} {2018})}\BibitemShut {NoStop}%
\bibitem [{\citenamefont {Liu}\ \emph {et~al.}(2008)\citenamefont {Liu},
  \citenamefont {Goree},\ and\ \citenamefont {Feng}}]{LiuPRE78}%
  \BibitemOpen
  \bibfield  {author} {\bibinfo {author} {\bibfnamefont {B.}~\bibnamefont
  {Liu}}, \bibinfo {author} {\bibfnamefont {J.}~\bibnamefont {Goree}}, \ and\
  \bibinfo {author} {\bibfnamefont {Y.}~\bibnamefont {Feng}},\ }\href {\doibase
  10.1103/PhysRevE.78.046403} {\bibfield  {journal} {\bibinfo  {journal} {Phys.
  Rev. E}\ }\textbf {\bibinfo {volume} {78}},\ \bibinfo {pages} {046403}
  (\bibinfo {year} {2008})}\BibitemShut {NoStop}%
\bibitem [{\citenamefont {Thomas}\ \emph {et~al.}(2004)\citenamefont {Thomas},
  \citenamefont {Williams},\ and\ \citenamefont {Silver}}]{stereopiv}%
  \BibitemOpen
  \bibfield  {author} {\bibinfo {author} {\bibfnamefont {E.}~\bibnamefont
  {Thomas}}, \bibinfo {author} {\bibfnamefont {J.~D.}\ \bibnamefont
  {Williams}}, \ and\ \bibinfo {author} {\bibfnamefont {J.}~\bibnamefont
  {Silver}},\ }\href {\doibase 10.1063/1.1755705} {\bibfield  {journal}
  {\bibinfo  {journal} {Physics of Plasmas}\ }\textbf {\bibinfo {volume}
  {11}},\ \bibinfo {pages} {L37} (\bibinfo {year} {2004})},\ \Eprint
  {http://arxiv.org/abs/https://doi.org/10.1063/1.1755705}
  {https://doi.org/10.1063/1.1755705} \BibitemShut {NoStop}%
\bibitem [{\citenamefont {Williams}(2011)}]{tomopiv}%
  \BibitemOpen
  \bibfield  {author} {\bibinfo {author} {\bibfnamefont {J.~D.}\ \bibnamefont
  {Williams}},\ }\href {\doibase 10.1063/1.3587090} {\bibfield  {journal}
  {\bibinfo  {journal} {Physics of Plasmas}\ }\textbf {\bibinfo {volume}
  {18}},\ \bibinfo {pages} {050702} (\bibinfo {year} {2011})},\ \Eprint
  {http://arxiv.org/abs/https://doi.org/10.1063/1.3587090}
  {https://doi.org/10.1063/1.3587090} \BibitemShut {NoStop}%
\bibitem [{\citenamefont {Han}\ and\ \citenamefont
  {Grier}(2005)}]{confTempSuspension}%
  \BibitemOpen
  \bibfield  {author} {\bibinfo {author} {\bibfnamefont {Y.}~\bibnamefont
  {Han}}\ and\ \bibinfo {author} {\bibfnamefont {D.~G.}\ \bibnamefont
  {Grier}},\ }\href {\doibase 10.1063/1.1844351} {\bibfield  {journal}
  {\bibinfo  {journal} {The Journal of Chemical Physics}\ }\textbf {\bibinfo
  {volume} {122}},\ \bibinfo {pages} {064907} (\bibinfo {year} {2005})},\
  \Eprint {http://arxiv.org/abs/https://doi.org/10.1063/1.1844351}
  {https://doi.org/10.1063/1.1844351} \BibitemShut {NoStop}%
\bibitem [{\citenamefont {Jackson}\ \emph {et~al.}(2016)\citenamefont
  {Jackson}, \citenamefont {Rubi},\ and\ \citenamefont {Bresme}}]{fluid2}%
  \BibitemOpen
  \bibfield  {author} {\bibinfo {author} {\bibfnamefont {N.}~\bibnamefont
  {Jackson}}, \bibinfo {author} {\bibfnamefont {J.~M.}\ \bibnamefont {Rubi}}, \
  and\ \bibinfo {author} {\bibfnamefont {F.}~\bibnamefont {Bresme}},\ }\href
  {\doibase 10.1080/08927022.2016.1168926} {\bibfield  {journal} {\bibinfo
  {journal} {Molecular Simulation}\ }\textbf {\bibinfo {volume} {42}},\
  \bibinfo {pages} {1214} (\bibinfo {year} {2016})},\ \Eprint
  {http://arxiv.org/abs/https://doi.org/10.1080/08927022.2016.1168926}
  {https://doi.org/10.1080/08927022.2016.1168926} \BibitemShut {NoStop}%
\bibitem [{\citenamefont {Schmidt}\ and\ \citenamefont
  {Piel}(2016)}]{SchmidtPOP23}%
  \BibitemOpen
  \bibfield  {author} {\bibinfo {author} {\bibfnamefont {C.}~\bibnamefont
  {Schmidt}}\ and\ \bibinfo {author} {\bibfnamefont {A.}~\bibnamefont {Piel}},\
  }\href {\doibase 10.1063/1.4960320} {\bibfield  {journal} {\bibinfo
  {journal} {Physics of Plasmas}\ }\textbf {\bibinfo {volume} {23}},\ \bibinfo
  {pages} {083704} (\bibinfo {year} {2016})}\BibitemShut {NoStop}%
\bibitem [{\citenamefont {Jepps}\ \emph {et~al.}(2000)\citenamefont {Jepps},
  \citenamefont {Ayton},\ and\ \citenamefont {Evans}}]{theo_CT1}%
  \BibitemOpen
  \bibfield  {author} {\bibinfo {author} {\bibfnamefont {O.~G.}\ \bibnamefont
  {Jepps}}, \bibinfo {author} {\bibfnamefont {G.}~\bibnamefont {Ayton}}, \ and\
  \bibinfo {author} {\bibfnamefont {D.~J.}\ \bibnamefont {Evans}},\ }\href
  {\doibase 10.1103/PhysRevE.62.4757} {\bibfield  {journal} {\bibinfo
  {journal} {Phys. Rev. E}\ }\textbf {\bibinfo {volume} {62}},\ \bibinfo
  {pages} {4757} (\bibinfo {year} {2000})}\BibitemShut {NoStop}%
\bibitem [{\citenamefont {Rickayzen}\ and\ \citenamefont
  {Powles}(2001)}]{theo_CT2}%
  \BibitemOpen
  \bibfield  {author} {\bibinfo {author} {\bibfnamefont {G.}~\bibnamefont
  {Rickayzen}}\ and\ \bibinfo {author} {\bibfnamefont {J.~G.}\ \bibnamefont
  {Powles}},\ }\href {\doibase 10.1063/1.1348024} {\bibfield  {journal}
  {\bibinfo  {journal} {The Journal of Chemical Physics}\ }\textbf {\bibinfo
  {volume} {114}},\ \bibinfo {pages} {4333} (\bibinfo {year} {2001})},\ \Eprint
  {http://arxiv.org/abs/https://doi.org/10.1063/1.1348024}
  {https://doi.org/10.1063/1.1348024} \BibitemShut {NoStop}%
\bibitem [{\citenamefont {Travis}\ and\ \citenamefont {Braga}(2006)}]{CTmd}%
  \BibitemOpen
  \bibfield  {author} {\bibinfo {author} {\bibfnamefont {K.~P.}\ \bibnamefont
  {Travis}}\ and\ \bibinfo {author} {\bibfnamefont {C.}~\bibnamefont {Braga}},\
  }\href {\doibase 10.1080/00268970601014880} {\bibfield  {journal} {\bibinfo
  {journal} {Molecular Physics}\ }\textbf {\bibinfo {volume} {104}},\ \bibinfo
  {pages} {3735} (\bibinfo {year} {2006})},\ \Eprint
  {http://arxiv.org/abs/https://doi.org/10.1080/00268970601014880}
  {https://doi.org/10.1080/00268970601014880} \BibitemShut {NoStop}%
\bibitem [{\citenamefont {Konopka}\ \emph {et~al.}(2000)\citenamefont
  {Konopka}, \citenamefont {Morfill},\ and\ \citenamefont
  {Ratke}}]{KonopkaPRL84}%
  \BibitemOpen
  \bibfield  {author} {\bibinfo {author} {\bibfnamefont {U.}~\bibnamefont
  {Konopka}}, \bibinfo {author} {\bibfnamefont {G.~E.}\ \bibnamefont
  {Morfill}}, \ and\ \bibinfo {author} {\bibfnamefont {L.}~\bibnamefont
  {Ratke}},\ }\href {\doibase 10.1103/PhysRevLett.84.891} {\bibfield  {journal}
  {\bibinfo  {journal} {Phys. Rev. Lett.}\ }\textbf {\bibinfo {volume} {84}},\
  \bibinfo {pages} {891} (\bibinfo {year} {2000})}\BibitemShut {NoStop}%
\bibitem [{\citenamefont {Wang}\ \emph {et~al.}(2001)\citenamefont {Wang},
  \citenamefont {Bhattacharjee},\ and\ \citenamefont {Hu}}]{WangPRL86}%
  \BibitemOpen
  \bibfield  {author} {\bibinfo {author} {\bibfnamefont {X.}~\bibnamefont
  {Wang}}, \bibinfo {author} {\bibfnamefont {A.}~\bibnamefont {Bhattacharjee}},
  \ and\ \bibinfo {author} {\bibfnamefont {S.}~\bibnamefont {Hu}},\ }\href
  {\doibase 10.1103/PhysRevLett.86.2569} {\bibfield  {journal} {\bibinfo
  {journal} {Phys. Rev. Lett.}\ }\textbf {\bibinfo {volume} {86}},\ \bibinfo
  {pages} {2569} (\bibinfo {year} {2001})}\BibitemShut {NoStop}%
\bibitem [{\citenamefont {Lai}\ and\ \citenamefont {I}(1999)}]{LaiPRE60}%
  \BibitemOpen
  \bibfield  {author} {\bibinfo {author} {\bibfnamefont {Y.-J.}\ \bibnamefont
  {Lai}}\ and\ \bibinfo {author} {\bibfnamefont {L.}~\bibnamefont {I}},\ }\href
  {\doibase 10.1103/PhysRevE.60.4743} {\bibfield  {journal} {\bibinfo
  {journal} {Phys. Rev. E}\ }\textbf {\bibinfo {volume} {60}},\ \bibinfo
  {pages} {4743} (\bibinfo {year} {1999})}\BibitemShut {NoStop}%
\bibitem [{\citenamefont {Hutchinson}(2012)}]{wakePRE}%
  \BibitemOpen
  \bibfield  {author} {\bibinfo {author} {\bibfnamefont {I.~H.}\ \bibnamefont
  {Hutchinson}},\ }\href {\doibase 10.1103/PhysRevE.85.066409} {\bibfield
  {journal} {\bibinfo  {journal} {Phys. Rev. E}\ }\textbf {\bibinfo {volume}
  {85}},\ \bibinfo {pages} {066409} (\bibinfo {year} {2012})}\BibitemShut
  {NoStop}%
\bibitem [{\citenamefont {Melzer}(2003{\natexlab{a}})}]{Melzer03}%
  \BibitemOpen
  \bibfield  {author} {\bibinfo {author} {\bibfnamefont {A.}~\bibnamefont
  {Melzer}},\ }\href@noop {} {\bibfield  {journal} {\bibinfo  {journal} {Phys.
  Rev. E}\ }\textbf {\bibinfo {volume} {67}},\ \bibinfo {pages} {016411}
  (\bibinfo {year} {2003}{\natexlab{a}})}\BibitemShut {NoStop}%
\bibitem [{\citenamefont {Schweigert}\ and\ \citenamefont
  {Peeters}(1995)}]{Schweigert95c}%
  \BibitemOpen
  \bibfield  {author} {\bibinfo {author} {\bibfnamefont {V.~A.}\ \bibnamefont
  {Schweigert}}\ and\ \bibinfo {author} {\bibfnamefont {F. M.}~\bibnamefont
  {Peeters}},\ }\href@noop {} {\bibfield  {journal} {\bibinfo  {journal} {Phys.
  Rev. B}\ }\textbf {\bibinfo {volume} {51}},\ \bibinfo {pages} {7700}
  (\bibinfo {year} {1995})}\BibitemShut {NoStop}%
\bibitem [{\citenamefont {Nelissen}\ \emph {et~al.}(2006)\citenamefont
  {Nelissen}, \citenamefont {Matulis}, \citenamefont {Partoens}, \citenamefont
  {Kong},\ and\ \citenamefont {Peeters}}]{Nelissen06}%
  \BibitemOpen
  \bibfield  {author} {\bibinfo {author} {\bibfnamefont {K.}~\bibnamefont
  {Nelissen}}, \bibinfo {author} {\bibfnamefont {A.}~\bibnamefont {Matulis}},
  \bibinfo {author} {\bibfnamefont {B.}~\bibnamefont {Partoens}}, \bibinfo
  {author} {\bibfnamefont {M.}~\bibnamefont {Kong}}, \ and\ \bibinfo {author}
  {\bibfnamefont {F.~M.}\ \bibnamefont {Peeters}},\ }\href@noop {} {\bibfield
  {journal} {\bibinfo  {journal} {Phys. Rev. E}\ }\textbf {\bibinfo {volume}
  {73}},\ \bibinfo {pages} {016607} (\bibinfo {year} {2006})}\BibitemShut
  {NoStop}%
\bibitem [{\citenamefont {Juan}\ \emph {et~al.}(1998)\citenamefont {Juan},
  \citenamefont {Huang}, \citenamefont {Hsu}, \citenamefont {Lai},\ and\
  \citenamefont {I}}]{JuanPRE58}%
  \BibitemOpen
  \bibfield  {author} {\bibinfo {author} {\bibfnamefont {W.-T.}\ \bibnamefont
  {Juan}}, \bibinfo {author} {\bibfnamefont {Z.-H.}\ \bibnamefont {Huang}},
  \bibinfo {author} {\bibfnamefont {J.-W.}\ \bibnamefont {Hsu}}, \bibinfo
  {author} {\bibfnamefont {Y.-J.}\ \bibnamefont {Lai}}, \ and\ \bibinfo
  {author} {\bibfnamefont {L.}~\bibnamefont {I}},\ }\href {\doibase
  10.1103/PhysRevE.58.R6947} {\bibfield  {journal} {\bibinfo  {journal} {Phys.
  Rev. E}\ }\textbf {\bibinfo {volume} {58}},\ \bibinfo {pages} {R6947}
  (\bibinfo {year} {1998})}\BibitemShut {NoStop}%
\bibitem [{\citenamefont {Klindworth}\ \emph {et~al.}(2000)\citenamefont
  {Klindworth}, \citenamefont {Melzer}, \citenamefont {Piel},\ and\
  \citenamefont {Schweigert}}]{KlindworthPRB61}%
  \BibitemOpen
  \bibfield  {author} {\bibinfo {author} {\bibfnamefont {M.}~\bibnamefont
  {Klindworth}}, \bibinfo {author} {\bibfnamefont {A.}~\bibnamefont {Melzer}},
  \bibinfo {author} {\bibfnamefont {A.}~\bibnamefont {Piel}}, \ and\ \bibinfo
  {author} {\bibfnamefont {V.~A.}\ \bibnamefont {Schweigert}},\ }\href
  {\doibase 10.1103/PhysRevB.61.8404} {\bibfield  {journal} {\bibinfo
  {journal} {Phys. Rev. B}\ }\textbf {\bibinfo {volume} {61}},\ \bibinfo
  {pages} {8404} (\bibinfo {year} {2000})}\BibitemShut {NoStop}%
\bibitem [{\citenamefont {Cheung}\ \emph {et~al.}(2003)\citenamefont {Cheung},
  \citenamefont {Samarian},\ and\ \citenamefont {James}}]{Cheung_2003}%
  \BibitemOpen
  \bibfield  {author} {\bibinfo {author} {\bibfnamefont {F.}~\bibnamefont
  {Cheung}}, \bibinfo {author} {\bibfnamefont {A.}~\bibnamefont {Samarian}}, \
  and\ \bibinfo {author} {\bibfnamefont {B.}~\bibnamefont {James}},\ }\href
  {\doibase 10.1088/1367-2630/5/1/375} {\bibfield  {journal} {\bibinfo
  {journal} {New Journal of Physics}\ }\textbf {\bibinfo {volume} {5}},\
  \bibinfo {pages} {75} (\bibinfo {year} {2003})}\BibitemShut {NoStop}%
\bibitem [{\citenamefont {Crocker}\ and\ \citenamefont
  {Grier}(1996)}]{crocker}%
  \BibitemOpen
  \bibfield  {author} {\bibinfo {author} {\bibfnamefont {J.~C.}\ \bibnamefont
  {Crocker}}\ and\ \bibinfo {author} {\bibfnamefont {D.~G.}\ \bibnamefont
  {Grier}},\ }\href {\doibase https://doi.org/10.1006/jcis.1996.0217}
  {\bibfield  {journal} {\bibinfo  {journal} {Journal of Colloid and Interface
  Science}\ }\textbf {\bibinfo {volume} {179}},\ \bibinfo {pages} {298 }
  (\bibinfo {year} {1996})}\BibitemShut {NoStop}%
\bibitem [{\citenamefont {Feng}\ \emph {et~al.}(2007)\citenamefont {Feng},
  \citenamefont {Goree},\ and\ \citenamefont {Liu}}]{feng}%
  \BibitemOpen
  \bibfield  {author} {\bibinfo {author} {\bibfnamefont {Y.}~\bibnamefont
  {Feng}}, \bibinfo {author} {\bibfnamefont {J.}~\bibnamefont {Goree}}, \ and\
  \bibinfo {author} {\bibfnamefont {B.}~\bibnamefont {Liu}},\ }\href {\doibase
  10.1063/1.2735920} {\bibfield  {journal} {\bibinfo  {journal} {Review of
  Scientific Instruments}\ }\textbf {\bibinfo {volume} {78}},\ \bibinfo {pages}
  {053704} (\bibinfo {year} {2007})},\ \Eprint
  {http://arxiv.org/abs/https://doi.org/10.1063/1.2735920}
  {https://doi.org/10.1063/1.2735920} \BibitemShut {NoStop}%
\bibitem [{\citenamefont {Ivanov}\ and\ \citenamefont
  {Melzer}(2007)}]{ivanov_heating}%
  \BibitemOpen
  \bibfield  {author} {\bibinfo {author} {\bibfnamefont {Y.}~\bibnamefont
  {Ivanov}}\ and\ \bibinfo {author} {\bibfnamefont {A.}~\bibnamefont
  {Melzer}},\ }\href {\doibase 10.1063/1.2714050} {\bibfield  {journal}
  {\bibinfo  {journal} {Review of Scientific Instruments}\ }\textbf {\bibinfo
  {volume} {78}},\ \bibinfo {pages} {033506} (\bibinfo {year} {2007})},\
  \Eprint {http://arxiv.org/abs/https://doi.org/10.1063/1.2714050}
  {https://doi.org/10.1063/1.2714050} \BibitemShut {NoStop}%
\bibitem [{\citenamefont {Menzel}\ \emph {et~al.}(2010)\citenamefont {Menzel},
  \citenamefont {Arp},\ and\ \citenamefont {Piel}}]{3D_MenzelPRL}%
  \BibitemOpen
  \bibfield  {author} {\bibinfo {author} {\bibfnamefont {K.~O.}\ \bibnamefont
  {Menzel}}, \bibinfo {author} {\bibfnamefont {O.}~\bibnamefont {Arp}}, \ and\
  \bibinfo {author} {\bibfnamefont {A.}~\bibnamefont {Piel}},\ }\href {\doibase
  10.1103/PhysRevLett.104.235002} {\bibfield  {journal} {\bibinfo  {journal}
  {Phys. Rev. Lett.}\ }\textbf {\bibinfo {volume} {104}},\ \bibinfo {pages}
  {235002} (\bibinfo {year} {2010})}\BibitemShut {NoStop}%
\bibitem [{\citenamefont {Himpel}\ \emph {et~al.}(2018)\citenamefont {Himpel},
  \citenamefont {Schütt}, \citenamefont {Miloch},\ and\ \citenamefont
  {Melzer}}]{himpel18}%
  \BibitemOpen
  \bibfield  {author} {\bibinfo {author} {\bibfnamefont {M.}~\bibnamefont
  {Himpel}}, \bibinfo {author} {\bibfnamefont {S.}~\bibnamefont {Schütt}},
  \bibinfo {author} {\bibfnamefont {W.~J.}\ \bibnamefont {Miloch}}, \ and\
  \bibinfo {author} {\bibfnamefont {A.}~\bibnamefont {Melzer}},\ }\href
  {\doibase 10.1063/1.5046049} {\bibfield  {journal} {\bibinfo  {journal}
  {Physics of Plasmas}\ }\textbf {\bibinfo {volume} {25}},\ \bibinfo {pages}
  {083707} (\bibinfo {year} {2018})},\ \Eprint
  {http://arxiv.org/abs/https://doi.org/10.1063/1.5046049}
  {https://doi.org/10.1063/1.5046049} \BibitemShut {NoStop}%
\bibitem [{\citenamefont {Thomas}\ \emph {et~al.}(2001)\citenamefont {Thomas},
  \citenamefont {Goldbeck}, \citenamefont {Hagl}, \citenamefont {Ivlev},
  \citenamefont {Konopka}, \citenamefont {Morfill}, \citenamefont {Rothermel},
  \citenamefont {S?tterlin},\ and\ \citenamefont {Zuzic}}]{Thomas_PFC2001}%
  \BibitemOpen
  \bibfield  {author} {\bibinfo {author} {\bibfnamefont {H.~M.}\ \bibnamefont
  {Thomas}}, \bibinfo {author} {\bibfnamefont {D.~D.}\ \bibnamefont
  {Goldbeck}}, \bibinfo {author} {\bibfnamefont {T.}~\bibnamefont {Hagl}},
  \bibinfo {author} {\bibfnamefont {A.~V.}\ \bibnamefont {Ivlev}}, \bibinfo
  {author} {\bibfnamefont {U.}~\bibnamefont {Konopka}}, \bibinfo {author}
  {\bibfnamefont {G.~E.}\ \bibnamefont {Morfill}}, \bibinfo {author}
  {\bibfnamefont {H.}~\bibnamefont {Rothermel}}, \bibinfo {author}
  {\bibfnamefont {R.}~\bibnamefont {S?tterlin}}, \ and\ \bibinfo {author}
  {\bibfnamefont {M.}~\bibnamefont {Zuzic}},\ }\href {\doibase
  10.1238/physica.topical.089a00016} {\bibfield  {journal} {\bibinfo  {journal}
  {Physica Scripta}\ }\textbf {\bibinfo {volume} {T89}},\ \bibinfo {pages} {16}
  (\bibinfo {year} {2001})}\BibitemShut {NoStop}%
\bibitem [{\citenamefont {Allen}\ and\ \citenamefont
  {Tildesley}(1987)}]{simu_book}%
  \BibitemOpen
  \bibfield  {author} {\bibinfo {author} {\bibfnamefont {M.~P.}\ \bibnamefont
  {Allen}}\ and\ \bibinfo {author} {\bibfnamefont {a.}~\bibnamefont
  {Tildesley}, \bibfnamefont {D.~J.}},\ }\href@noop {} {\emph {\bibinfo {title}
  {Computer Simulation of Liquids}}}\ (\bibinfo  {publisher} {Clarendon Press, New York/Oxford University Press, Oxford},\ \bibinfo {year}
  {1987})\BibitemShut {NoStop}%
\bibitem [{\citenamefont {Homann}\ \emph {et~al.}(1997)\citenamefont {Homann},
  \citenamefont {Melzer}, \citenamefont {Peters},\ and\ \citenamefont
  {Piel}}]{HomannPRE56}%
  \BibitemOpen
  \bibfield  {author} {\bibinfo {author} {\bibfnamefont {A.}~\bibnamefont
  {Homann}}, \bibinfo {author} {\bibfnamefont {A.}~\bibnamefont {Melzer}},
  \bibinfo {author} {\bibfnamefont {S.}~\bibnamefont {Peters}}, \ and\ \bibinfo
  {author} {\bibfnamefont {A.}~\bibnamefont {Piel}},\ }\href {\doibase
  10.1103/PhysRevE.56.7138} {\bibfield  {journal} {\bibinfo  {journal} {Phys.
  Rev. E}\ }\textbf {\bibinfo {volume} {56}},\ \bibinfo {pages} {7138}
  (\bibinfo {year} {1997})}\BibitemShut {NoStop}%
\bibitem [{\citenamefont {Liu}\ and\ \citenamefont {Goree}(2005)}]{LiuPRE71}%
  \BibitemOpen
  \bibfield  {author} {\bibinfo {author} {\bibfnamefont {B.}~\bibnamefont
  {Liu}}\ and\ \bibinfo {author} {\bibfnamefont {J.}~\bibnamefont {Goree}},\
  }\href {\doibase 10.1103/PhysRevE.71.046410} {\bibfield  {journal} {\bibinfo
  {journal} {Phys. Rev. E}\ }\textbf {\bibinfo {volume} {71}},\ \bibinfo
  {pages} {046410} (\bibinfo {year} {2005})}\BibitemShut {NoStop}%
\bibitem [{\citenamefont {Homann}\ \emph {et~al.}(1998)\citenamefont {Homann},
  \citenamefont {Melzer}, \citenamefont {Peters}, \citenamefont {Madani},\ and\
  \citenamefont {Piel}}]{HomanPLA}%
  \BibitemOpen
  \bibfield  {author} {\bibinfo {author} {\bibfnamefont {A.}~\bibnamefont
  {Homann}}, \bibinfo {author} {\bibfnamefont {A.}~\bibnamefont {Melzer}},
  \bibinfo {author} {\bibfnamefont {S.}~\bibnamefont {Peters}}, \bibinfo
  {author} {\bibfnamefont {R.}~\bibnamefont {Madani}}, \ and\ \bibinfo {author}
  {\bibfnamefont {A.}~\bibnamefont {Piel}},\ }\href {\doibase
  https://doi.org/10.1016/S0375-9601(98)00141-8} {\bibfield  {journal}
  {\bibinfo  {journal} {Physics Letters A}\ }\textbf {\bibinfo {volume}
  {242}},\ \bibinfo {pages} {173 } (\bibinfo {year} {1998})}\BibitemShut
  {NoStop}%
\bibitem [{\citenamefont {Nunomura}\ \emph
  {et~al.}(2002{\natexlab{b}})\citenamefont {Nunomura}, \citenamefont {Goree},
  \citenamefont {Hu}, \citenamefont {Wang},\ and\ \citenamefont
  {Bhattacharjee}}]{nunomura}%
  \BibitemOpen
  \bibfield  {author} {\bibinfo {author} {\bibfnamefont {S.}~\bibnamefont
  {Nunomura}}, \bibinfo {author} {\bibfnamefont {J.}~\bibnamefont {Goree}},
  \bibinfo {author} {\bibfnamefont {S.}~\bibnamefont {Hu}}, \bibinfo {author}
  {\bibfnamefont {X.}~\bibnamefont {Wang}}, \ and\ \bibinfo {author}
  {\bibfnamefont {A.}~\bibnamefont {Bhattacharjee}},\ }\href {\doibase
  10.1103/PhysRevE.65.066402} {\bibfield  {journal} {\bibinfo  {journal} {Phys.
  Rev. E}\ }\textbf {\bibinfo {volume} {65}},\ \bibinfo {pages} {066402}
  (\bibinfo {year} {2002}{\natexlab{b}})}\BibitemShut {NoStop}%
\bibitem [{\citenamefont {Khrapak}\ \emph {et~al.}(2005)\citenamefont
  {Khrapak}, \citenamefont {Ratynskaia}, \citenamefont {Zobnin}, \citenamefont
  {Usachev}, \citenamefont {Yaroshenko}, \citenamefont {Thoma}, \citenamefont
  {Kretschmer}, \citenamefont {H\"ofner}, \citenamefont {Morfill},
  \citenamefont {Petrov},\ and\ \citenamefont {Fortov}}]{khrapak}%
  \BibitemOpen
  \bibfield  {author} {\bibinfo {author} {\bibfnamefont {S.~A.}\ \bibnamefont
  {Khrapak}}, \bibinfo {author} {\bibfnamefont {S.~V.}\ \bibnamefont
  {Ratynskaia}}, \bibinfo {author} {\bibfnamefont {A.~V.}\ \bibnamefont
  {Zobnin}}, \bibinfo {author} {\bibfnamefont {A.~D.}\ \bibnamefont {Usachev}},
  \bibinfo {author} {\bibfnamefont {V.~V.}\ \bibnamefont {Yaroshenko}},
  \bibinfo {author} {\bibfnamefont {M.~H.}\ \bibnamefont {Thoma}}, \bibinfo
  {author} {\bibfnamefont {M.}~\bibnamefont {Kretschmer}}, \bibinfo {author}
  {\bibfnamefont {H.}~\bibnamefont {H\"ofner}}, \bibinfo {author}
  {\bibfnamefont {G.~E.}\ \bibnamefont {Morfill}}, \bibinfo {author}
  {\bibfnamefont {O.~F.}\ \bibnamefont {Petrov}}, \ and\ \bibinfo {author}
  {\bibfnamefont {V.~E.}\ \bibnamefont {Fortov}},\ }\href {\doibase
  10.1103/PhysRevE.72.016406} {\bibfield  {journal} {\bibinfo  {journal} {Phys.
  Rev. E}\ }\textbf {\bibinfo {volume} {72}},\ \bibinfo {pages} {016406}
  (\bibinfo {year} {2005})}\BibitemShut {NoStop}%
\bibitem [{\citenamefont {Akdim}\ and\ \citenamefont
  {Goedheer}(2003)}]{akdim_plasma}%
  \BibitemOpen
  \bibfield  {author} {\bibinfo {author} {\bibfnamefont {M.~R.}\ \bibnamefont
  {Akdim}}\ and\ \bibinfo {author} {\bibfnamefont {W.~J.}\ \bibnamefont
  {Goedheer}},\ }\href {\doibase 10.1103/PhysRevE.67.066407} {\bibfield
  {journal} {\bibinfo  {journal} {Phys. Rev. E}\ }\textbf {\bibinfo {volume}
  {67}},\ \bibinfo {pages} {066407} (\bibinfo {year} {2003})}\BibitemShut
  {NoStop}%
\bibitem [{\citenamefont {Havnes}\ \emph {et~al.}(1987)\citenamefont {Havnes},
  \citenamefont {Goertz}, \citenamefont {Morfill}, \citenamefont {Grün},\ and\
  \citenamefont {Ip}}]{havnes}%
  \BibitemOpen
  \bibfield  {author} {\bibinfo {author} {\bibfnamefont {O.}~\bibnamefont
  {Havnes}}, \bibinfo {author} {\bibfnamefont {C.~K.}\ \bibnamefont {Goertz}},
  \bibinfo {author} {\bibfnamefont {G.~E.}\ \bibnamefont {Morfill}}, \bibinfo
  {author} {\bibfnamefont {E.}~\bibnamefont {Gr"{u}n}}, \ and\ \bibinfo {author}
  {\bibfnamefont {W.}~\bibnamefont {Ip}},\ }\href {\doibase
  10.1029/JA092iA03p02281} {\bibfield  {journal} {\bibinfo  {journal} {Journal
  of Geophysical Research: Space Physics}\ }\textbf {\bibinfo {volume} {92}},\
  \bibinfo {pages} {2281} (\bibinfo {year} {1987})}\BibitemShut {NoStop}
\bibitem [{\citenamefont {Goertz}\ \emph {et~al.}(2011)\citenamefont {Goertz},
  \citenamefont {Greiner},\ and\ \citenamefont {Piel}}]{Goertz11}%
  \BibitemOpen
  \bibfield  {author} {\bibinfo {author} {\bibfnamefont {I.}~\bibnamefont
  {Goertz}}, \bibinfo {author} {\bibfnamefont {F.}~\bibnamefont {Greiner}}, \
  and\ \bibinfo {author} {\bibfnamefont {A.}~\bibnamefont {Piel}},\ }\href@noop
  {} {\bibfield  {journal} {\bibinfo  {journal} {Physics of Plasmas}\ }\textbf
  {\bibinfo {volume} {18}},\ \bibinfo {pages} {013703} (\bibinfo {year}
  {2011})}\BibitemShut {NoStop}%
\end{thebibliography}
\end{document}